\documentclass[pre,superscriptaddress,preprint]{revtex4}
\usepackage{amsmath,amssymb,amsfonts}
\usepackage{graphicx}
\begin{document}
\title{Similarities between GSH, Hypoplasticity and KCR}
\author{Yimin Jiang} 
\affiliation{Central South University, Changsha 410083, China}
\author{Mario Liu}
\affiliation{Theoretische Physik, Universität T\"{u}bingen,72076
T\"{u}bingen, Germany}
\begin{abstract}
Accounting for elasto-plastic motion in granular media, hypoplasticity is a state-of-the-art constitutive model derived from data accumulated over many decades.  In contrast, GSH, a hydrodynamic theory, is derived from general principles of physics, 
with comparatively few inputs from experiments, yet sporting an applicability ranging from static stress distribution via elasto-plastic motion to fast dense flow, including non-uniform ones such as a shear band. Comparing both theories, we find great similarities for uniform, slow, elasto-plastic motion. We also find that proportional paths and the Goldscheider rule used to construct barodesy, another, more recent  constitutive model, are natural  results of GSH's equations. This is useful as it gives these constitutive relations a solid foundation in physics, and in reverse, GSH a robust connection to reality. The same symbiotic relation exists between GSH and KCR, or  Kamrin's non-local constitutive relation, a model that was successfully employed to account for a wide shear band in  split bottom cells.  

\end{abstract}

\maketitle 
\tableofcontents
\vspace{2cm}
\noindent
{\bf Notations:}\\

\begin{tabular}{| l | r| }\hline
 $v_{ij}$ & $\equiv\frac12(\nabla_iv_j+\nabla_jv_i)$, the strain rate, \\ \hline
 $u_{ij}$ &  $\equiv\varepsilon^{elast}_{ij}$,\qquad\quad\,\, the elastic strain, \\ \hline
 $\pi_{ij}$ &the elastic stress,\\ \hline
$\sigma_{ij}$ & the Cauchy stress, \\ \hline
$x^*_{ij}$ & the traceless part of $x_{ij}$,  \\ \hline
\end{tabular}

\begin{tabular}{| l |c| r| }\hline
$\Delta\equiv -u_{\ell\ell}$, &$P_\Delta\equiv\pi_{\ell\ell}/3$, &
$P\equiv\sigma_{\ell\ell}/3$,  \\ \hline
$v_s\equiv\sqrt{v^*_{ij}v^*_{ij}}$ & $\equiv||v^*_{ij}||$,\qquad $$ &
$$  \\ \hline
$u_s\equiv\sqrt{ u^*_{ij}u^*_{ij}}$, & $\pi_s\equiv\sqrt{ \pi^*_{ij}\pi^*_{ij}}$, & $ \sigma_s\equiv\sqrt{ \sigma^*_{ij}\sigma^*_{ij}}$. \\ \hline
\end{tabular}

\section{Introduction\label{intro}} 
Being a subject of practical importance, elasto-plastic deformation of dense granular media has been under the focus of engineering research for many decades if not
a century~\cite{schofield,nedderman,wood1990,kolymbas1,kolymbas2,gudehus2010}. The state of geotechnical theories, however, remains confusing for outsiders: Innumerable models compete, employing strikingly different
expressions. In a recent book on soil mechanics, Gudehus employed  metaphors such as {\em morass of equations} and {\em jungle of data} to describe the situation~\cite{gudehus2010}. 

Granular dynamics is frequently modeled employing  the strategy of
{\em rational mechanics}, by postulating a function $\mathfrak{C}_{ij}$,  such that the constitutive relation, 
\begin{equation}
{\partial}_{t}\sigma _{ij}=\mathfrak{C}_{ij}(\sigma _{ij},v_{k\ell},\rho)
\end{equation}•
holds, whereby  $\mathfrak{C}_{ij}$ is a function of the Cauchy stress $\sigma _{ij}$, strain rate  $v_{k\ell}$ (the symmetric part of the velocity gradient) and density $\rho$.   (More generally, ${\partial}_{t}\equiv\frac\partial{\partial_t}$ is to be replaced by an  objective  derivative, such as Jauman -- here and below.)  

Together with mass and momentum conservation, it forms  a closed set of equations for $\sigma _{ij},\rho$ and the velocity $v_{i}$. Hypoplasticity~\cite{kolymbas1,kolymbas2} is such a theory, same as barodesy~\cite{barodesy}. 
Both use a single constitutive equation,  without the recourse to yield or potential surfaces. Although elasto-plastic theories are somewhat different, Einav~\cite{h^2} showed that the latter can be seen as a special case of the former. All these models  yield realistic accounts of the complex elasto-plastic motion, providing us with vast amount of highly condensed and intelligently organized empirical data. 
We concentrate on comparing GSH to hypoplasticity here, because elasto-plastic theories, with its case separations, are not easy to deal with analytically, and because  a comparison to barodesy already exists~\cite{GSH&Barodesy}. 

Typically, the {\em hypoplastic model} starts from the rate-independent constitutive relation, 
\begin{equation}\label{3b-1}
\partial_t\sigma_{ij}=H_{ijk\ell}v_{k\ell}+
\Lambda_{ij}\sqrt{v_{k\ell}v_{k\ell}+\epsilon v_{\ell\ell}^2}, \end{equation} with $H_{ijk\ell},\Lambda_{ij}$ and $\epsilon$ functions of stress and density. Though continually improving functional dependence is being proposed for them, some known and systematic shortcomings remain:
\begin{enumerate}
\item The applicability of hypoplasticity is confined to uniform systems. Lacking a characteristic length scale, it does not account for shear band or evanescent creep flow (ie. the exponentially decaying penetration of flows into the static region). 
There have been two main approaches to ameliorate this: introduction of gradient terms~\cite{wu2}, or addition of variables to account for the Crosserat rotation and   couple stress~\cite{wu1}.
The latter works well, but leads to a  rather more complex theory. And it provokes questions about the underlying physics: If couple stress is indeed important in the shear band, because it is fluid, why then is it not important in the uniformly fluid and gaseous state of granular media -- or, coming to think of it, in nematic liquid crystals~\cite{deGennes}? 

\item In hypoplasticity, rate-independence is an input. Yet it is not a robust feature of granular behavior. Fast dense flow is rate-dependent, as is the critical stress  in the presence of external perturbations (such as given by tapping). Rate-independence needs to be explained, understood, and the extent to where it is valid clarified. 

\item Elastic waves exist in granular media. Yet if one tries to account for them employing Eq.(\ref{3b-1}), one finds falsely that they should always be over-damped. 
More generally, granular media appear softer and more dissipative during elasto-plastic motion than in many other circumstances -- foremost elastic waves, but also  incremental stress-strain relations and ratcheting. Focusing on the former, hypoplasticity struggles to account for any of the latter. (See however works on intergranular strains~\cite{Herle}.) 

\item Fast dense flow and collisional flow are not accounted for. 

\item Finally, hypoplasticity lacks energetics. One does not know whether a given state is stable because it has an energy minimum, or instable because the energy turns concave. Nor do we know how much energy is being dissipated say in the critical state. 
Moreover, $H_{ijk\ell}$ has 36 elements, $\Lambda_{ij}$ six, all functions of a tensor and a scalar. These are a lot of functions to choose. So it would be helpful if  the second derivative of the energy were related to $H_{ijk\ell}$ and/or $\Lambda_{ij}$ (contradicting in fact the name {\it hypoplasticity}).

\end{enumerate}
{Granular solid hydrodynamics} (GSH) is derived from basic principles of physics, including conservation laws and the second law of thermodynamics,
by  employing {\it the hydrodynamic procedure}~\cite{Khal,LL6}. The result is a 
continuum-mechanical theory~\cite{granL3,granR2,granR3,granRgudehus,granR4,GGas,granRexp}  that leaves relatively little leeway, with a handful of scalar parameters to fit experiments. 
The most important one is the energy $w$, chosen such, first of all, that elastic waves, incremental stress-strain relation and static stress distributions are well accounted for. Second, $w$ contains the information that no stable static stress distribution exists if the density $\rho$  is too small for enduring contacts, or if the shear stress is too large for the given pressure. Third, by reducing GSH to Eq.(\ref{3b-1}) for slow, uniform shear rates,  $w$'s second derivative is found to yield $H_{ijk\ell}$ and $\Lambda_{ij}$. 

The thermodynamic consistency of hypoplasticity has been demonstrated independently by Fang et al, see \cite{Fang1,Fang2}. This was done generally, without specifying an expression for $w$.

The energy $w$ is taken as a function of the density $\rho$, granular temperature $T_g$ and elastic strain $u_{ij}$.
The state variable $u_{ij}$ accounts for {\it the slowly-varying, large-scaled elastic deformations of the grains, }
with the  elastic stress given by the hyper-elastic relation, 
\begin{equation}\label{u-stress}
\pi_{ij}\equiv-{\partial w}/{\partial u_{ij}}|_{\rho}.
\end{equation}

The state variable $T_g$ quantifies {\it the fast fluctuating elastic and kinetic energy of the grains}. It is frequently proportional to the shear rate, 
$T_g\propto v_s \,\,(\equiv\dot\gamma)$, and  a useful indicator for the system's behavior, from the quasi-static via the dense flow to the collisional one.

For collisional and fast dense flow, with a shear rate of 
$10^{3}$/s and more, there is  vigorous jiggling and aggitation, so $T_g$ is obviously important. 
For elasto-plastic motion (usually referred to as ``quasi-static" but see below), with shear rates around $10^{-3}$/s, $T_g$ may 
seem irrelevant: Aside from an occasional slip, grains only participate in 
the motion dictated by the macroscopic shear rate, with no perceptible deviations that may contribute to $T_g$. 
Nevertheless, such slips lead to vibrations in the region that imply a $T_g$ still many orders of magnitude larger than the true 
temperature of the grains. More importantly, as we shall see, these vibrations are what lead to plasticity.  

If $T_g\propto v_s$ is yet smaller, below $10^{-5}$/s, GSH predicts a rate-dependent transition, to irrelevance of $T_g$ and diminishing plasticity. This is supported by recent experiments~\cite{QuasiEla}. Here, the system becomes truly quasi-static, better, {\it quasi-elastic}, until statics reigns, $T_g\equiv0$. 

In what follows, we shall briefly present GSH in Sec.\ref{GSH}, first its three coupled relaxation equations and an expression for the elastic energy, followed by a discussion of how best to understand $T_g$ and $u_{ij}$, 
and why a second  $T_g$ for the fluctuating elastic energy alone is counter-productive. 
In Sec.\ref{hypoplastic motion}, we obtain the critical state as the steady-state solution of the three relaxation equations. The approach to the state, for both high and low initial densities, is  given by a simple analytic solution. In addition, GSH accounts for the variously  observed fact that disturbing the critical state with some tapping, its stress becomes highly rate-dependent.  
We then compare GSH in detail to hypoplasticity and neo-hypoplasticity, and discuss how the observed proportional paths and Goldscheider rule are natural results of GSH's equations. Finally, we compare four more recently proposed elastic energies, showing that all four appear to agree well, though their second derivatives differ close to yield.  

In Sec.\ref{aging}, we consider remifications of GSH if the stress is held constant, stressing the difference to rate-controlled experiments, explaining why a soft spring is useful for the latter, and a triaxial apparatus for the former. Then we calculate  creep, shear band, angle of repose  and stability (for the infinite, uniform geometry). Finally, we demonstrate the similarity of GSH to Kamrin's nonlocal KCR model.

In Sec.\ref{quasiEla}, we discuss the transition to the quasi-elastic regime for extremely slow rates.
\section{The Equations of GSH \label{GSH}    }

\subsection{The state variables and their dynamics}
A { complete set of state variables} is one that
uniquely determines a macroscopic state, including its energy and stress. GSH's  state variables are the density $\rho$, velocity $v_i$, the granular temperature $T_g$ and the elastic strain $u_{ij}$. 
The latter two quantify, respectively, the temporally fluctuating elastic and kinetic energy of the grains and their slowly-varying large-scaled deformations. $T_g$ is increased by viscous heating $(fv_s)^2$, decreased by relaxation, and its  balance equation is
\begin{equation}
\label{Tg2} 
\partial_tT_g=-R_T[T_g(1-\xi_T^2\nabla^2_i)T_g-(fv_s)^2-T_a^2],
\end{equation}
with $R_TT_g$ the relaxation rate, and $R_T\xi_T^2\nabla^2_iT_g$ accounting for $T_g$-diffusion.  (The strain rate is  $v_{ij}\equiv\frac12(\nabla_iv_j+\nabla_jv_i)$, its traceless part $v^*_{ij}$, and $v_s\equiv\sqrt{v^*_{ij}v^*_{ij}}>0$.)
$T_a$ is the ``ambient temperature" maintained by tapping or a sound field.
For a steady, uniform $v_s$, we have 
\begin{equation}\label{fVs}
T_g=\sqrt{ f^2 v_s^2+T_a^2},
\end{equation}
implying $T_g= f v_s$ if there is no external perturbation, and
$T_g=T_a$ if the shear rate is zero.  As is easy to see from Eqs.(\ref{dot u},\ref{dot u2}) below, it is  $T_g= f v_s$ that leads to rate-independence. The easiest way to destroy it is therefore  to introduce some $T_a$, see Sec.\ref{external perturbation}.

The  importance of $T_g$ in the collisional state is easy to accept. Recent progresses show it exists equally in the turbulent flows of rapid dry granular avalanches~\cite{Fang3}. That $T_g$ is relevant even in elasto-plastic motion may surprise, yet as already discussed in the introduction, this is a result of $T_g$,  many orders of magnitude larger than the true temperature, being the physics underlying plasticity, see the next few paragraphs.  

The elastic strain $u_{ij}$ alone suffices to account for elasticity and plasticity. To understand this, one needs two steps: First, granular stresses relax with a variable relaxation time. Second, this  relaxation is closely related to plasticity. 

The free surface of a granular system at rest is frequently tilted. When perturbed, when the grains jiggle and $T_g\not=0$, the tilted surface will decay and become horizontal. The stronger the grains jiggle
and slide, the faster the decay is. We take this as indicative of a system
that is {elastic for $T_g=0$, transiently elastic for $T_g\not=0$,
with a stress relaxation rate $\propto T_g$}. 
A relaxing stress is typical of any viscous-elastic or transiently elastic systems, such as polymers~\cite{polymer-1,polymer-2,polymer-3,polymer-4}. 
In granular media, the relaxation rate is not a material constant, but a function of the state variable $T_g$ -- a behavior that we name  {\em variable transient elasticity}. 

For given $\rho$,  the stress $\pi_{ij}\equiv-\partial w/\partial u_{ij}$ is a monotonic function of $u_{ij}$ (as long as the energy $w$ is stable and convex, $\partial^2 w/\partial u_{ij}\partial u_{k\ell}>0$). Therefore, there is no principle difference in considering either relaxation. Yet since strain is a geometric quantity, stress a physical one that includes material parameters such as the stiffness (stress dependent and highly anisotropic in granular media), considering  strain relaxation is simpler. 

We take the equation of motion for $u_{ij}$ such to reflect {\it variable transient elasticity}, finding that it suffices to capture elasto-plasticity. 
Starting from pure elasticity, $\partial_tu_{ij}-v_{ij}=0$, adding a relaxation term,  $\partial_tu_{ij}-v_{ij}=-u_{ij}/\tau$, we allow for two tensorial coefficients, $\lambda_{ijk\ell}$  and $\alpha_{ijk\ell}$, to parameterize the efficacy of relaxation and deformation rate, 
\begin{equation}\label{dot u}
\partial_tu_{ij}-v_{ij}+\alpha_{ijk\ell}\,v_{k\ell}=-(\lambda_{ijk\ell}T_g)\,u_{k\ell}. 
\end{equation} 
If stress anisotropy changes appreciably, convective terms (or the so-called objective time derivative) become important, then one needs the substitution (see the derivation in~\cite{polymer-1})
\begin{equation}
\partial_t u_{ij}\to(\partial_t+v_k\nabla_k)u_{ij}
+u_{ik}\nabla_{j}v_k+u_{jk}\nabla_{i}v_k.
\end{equation}
Since $\partial_t\pi_{mn}=(\partial\pi_{mn}/\partial u_{ij})\partial_tu_{ij}=-(\partial^2w/\partial u_{mn}\partial u_{ij})\partial_tu_{ij}$,  and $T_g=f v_s$, see Eq.(\ref{fVs}), this simple equation possesses the full hypoplastic structure, with 
\begin{align}\label{dot u2}
H_{mnk\ell}=(\partial\pi_{mn}/\partial u_{ij})(\delta_{ik}\delta_{j\ell}-\alpha_{ijk\ell}),\quad
\Lambda_{mn}=(\partial\pi_{mn}/\partial u_{ij})\lambda_{ijk\ell}u_{k\ell}.
\end{align}
Given the equality of structure, GSH is as capable of accounting for elastoplasticity as hypoplasticity, if an appropriate energy expression can be found. (Allowing the density to vary yields slightly more complicated expressions.)
Next, dividing $u_{ij}$ into its trace $\Delta\equiv-u_{\ell\ell}$ and traceless part $u_{ij}^*$, denoting
$u_s\equiv\sqrt{u^*_{ij}u^*_{ij}}>0,\quad \hat u_{ij}\equiv u_{ij}^*/u_s, \quad\hat v_{ij}\equiv v_{ij}^*/v_s$,
and specifying the two tensors with two elements each, $\alpha,\alpha_1,\lambda,\lambda_1$, we arrive at
\begin{align}
\label{2c-7}
\partial_t\Delta+(1-\alpha )v_{\ell\ell} -\alpha_1u^*_{ij}v^*_{ij}
=-\lambda_1T_g\Delta, 
\\\label{2c-8} 
\partial_tu^*_{ij}-(1-\alpha )v^*_{ij}
= -\lambda T_gu^*_{ij},
\end{align} 
which may equivalently be rewritten as two relaxation equations,
\begin{align}
\label{eqU}
\partial_t u_{ij}^*&=-\lambda T_g[u_{ij}^*-u_c\hat v_{ij}\frac{fv_s}{T_g}], \quad u_c\equiv\frac{1-\alpha}{\lambda f},\,\,\frac{\Delta_c}{u_c}\equiv\frac{\alpha_1}{\lambda_1f},
\\\label{eqD}
\partial_t\Delta&+(1-\alpha)v_{\ell\ell}=-\lambda_1T_g\left[\Delta-\Delta_c\hat v_{ij}\hat u_{ij}\frac{fv_s}{T_g}\frac{u_s}{u_c}\right].
\end{align} 
Eqs.(\ref{Tg2},\ref{eqU},\ref{eqD}) are the three relaxation equations of GSH. They contain three inverse time scales: $R_TT_g,\lambda T_g,\lambda_1 T_g$,  a length scale $\xi_T$, and the parameters $f,\alpha,\alpha_1$. 
In dense media,  $R_TT_g$ is large, of order $10^3$/s or more. With $T_g=fv_s$, the scales $\lambda T_g,\lambda_1T_g$ are, for the shear rates typical of soil-mechanical experiments, orders of magnitude smaller, around 1/s for $v_s=10^{-2}$/s.  The length scale $\xi_T$ is a few to a few tens granular diameters in dense media. 

The steady state solution: $T_g=fv_s$, $u_s=u_c$ and $\Delta=\Delta_c$, implying constant elastic strain and stress at given shear rate,  is clearly the critical state, see Sec.~\ref{critical state} below.

Conservation of momentum and mass, 
\begin{equation}\label{cons}
\partial_t(\rho v_i)+\nabla_j(\sigma_{ij}+\rho v_iv_j)=g_i\rho,\,\,\,\partial_t\rho=-\nabla_i(\rho v_i),
\end{equation}
close the set of equations, where the Cauchy stress $\sigma_{ij}$ is  
\begin{align}\label{sigma}
\sigma_{ij}=(1-\alpha)\pi_{ij}-\alpha_1u^*_{ij}P_\Delta+[P_T\delta_{ij}-\eta_1T_g v^*_{ij}]†, \quad P\equiv\sigma_{\ell\ell}/3,
\end{align}
see~\cite{granR2} for derivation. The (off-diagonal Onsager) coefficients $\alpha,\alpha_1$, a necessary consequence of the same coefficients  in Eqs.(\ref{2c-7},\ref{2c-8}), soften and mix the stress components. The term preceded by $\alpha_1$ is smaller by one order in the small elastic strain $u_{ij}$ and may usually be neglected (though not in Sec.~\ref{Hypoplasticity}).  $\alpha,\alpha_1$ are functions of the density for elasto-plastic motion, though they vanish identically  for static stresses, see~Eq.(\ref{alphaEla}). The pressure $P_T\propto T_g^2$ is exerted by jiggling grains, and the last term is the viscous stress.  Both are $\propto v_s^2$ for $T_g\propto v_s$ and relevant only for fast collisional and dense flows, giving rise to Bagnold scaling there. 

\subsection{The energy\label{granEn}} 

An explicit expression for the stress via Eqs.(\ref{u-stress},\ref{sigma}), is delivered by the energy $w$ that we take to be
$w=w_\Delta(\Delta,u_s,\rho)+w_T(T_g,\rho)$. The part relevant for elasto-plastic flow is $w_\Delta$. ($w_T$, important for collisional and dense flow, is neglected here.) Denoting 
\begin{equation}
\Delta\equiv -u_{\ell\ell},\quad P_\Delta\equiv\pi_{\ell\ell}/3, \quad u_s^2\equiv
u^*_{ij}u^*_{ij}, \quad\pi_s^2\equiv \pi^*_{ij}\pi^*_{ij},
\end{equation}
 with
$u^*_{ij},\pi^*_{ij}$ the respective traceless tensors, we take $w_\Delta$ as 
\begin{align}
\label{2b-2} 
w_\Delta&=\sqrt{\Delta }(2 {\mathcal B} \Delta^2/5+ {\mathcal A}u_s^2),
\quad ({\cal A,B}>0,)
\\\label{2b-2a} 
\pi_{ij}&=\sqrt\Delta({\cal B}\Delta+{\cal A}
{u_s^2}/{2\Delta})\delta _{ij}-2{\cal A}\sqrt\Delta\, u_{ij}^*, 
\\\label{2b-2b} 
P_\Delta&=\sqrt\Delta({\cal B}\Delta+{\cal A}
{u_s^2}/{2\Delta}),\quad \pi_s=2{\cal A}\sqrt\Delta\, u_s,
\\\label{2b-1}
\frac{4P_\Delta}{\pi_s}&=\frac{2\cal B}{\cal
A}\frac{\Delta}{u_s}+\frac{u_s}{\Delta},\quad 
\hat u_{ij}\equiv \frac{u_{ij}^*}{u_s}=-\frac{\pi_{ij}^*}{\pi_s}\equiv -\hat\pi_{ij}.
\end{align} 
Given by the energy of linear elasticity multiplied by $\sqrt\Delta$, this form is clearly inspired by the Hertzian contact, though its connection to granular elasticity goes far beyond that. 
It includes both {\it stress-induced anisotropy} and the {\it convexity transition} (see below). In addition, the expression for the elastic stress $\pi_{ij}$ has been validated for: 
\begin{itemize}
\item  Static stress distribution in three classic geometries: silo, sand pile,
point load on a granular sheet, calculated employing $\nabla_i\pi_{ij}=\rho g_i$, see~\cite{ge1,ge2,granR1}. 
\item  Incremental stress-strain relation at varying static stresses~\cite{KJ,AH,incre}.
\item  Propagation of anisotropic elastic waves at varying static stresses~\cite{jia2009,ge4}. 
\end{itemize}
The GSH framework is general enough that one may employ any energy expression one deems appropriate. In fact, the above expression has been chosen for simplicity and manipulative ease, with a more realistic one given in~\cite{3inv}. In  Sec.~\ref{constRel}, we shall show and compare a number of different elastic energy expressions for realistic modeling.

 {\it Stress-induced anisotropy}: In  linear elasticity, $w\propto u_s^2$, the velocity of  elastic waves $\propto\sqrt{\partial^2w/\partial u_s^2}\,$ does not depend on $u_s$, or equivalently, the stress. For any exponent other than two, the velocity depends on the stress, and is anisotropic if the stress is.  

{\it Convexity Transition}: 
There are values of stress and density in granular media for which elastic solutions do not exist:  
First, for densities smaller than the random loose packing density,  $\rho<\rho_{\ell p}$, at which the grains start to loose contact with one another, no static elastic states can be maintained. Second, static elastic states collapse when the shear stress is too large for the given pressure. 
On a tilted plane, the angle at which this happens  is frequently referred to as the angle of  stability, $\varphi_{st}$. The actually measured angle varies with the spacial dependence of the stress components and does not have a unique value. But in an infinite geometry,  for uniform stresses,  elastic solutions fail at a well-defined value. 

Note that  $\varphi_{st}$, accounting for collapses of static stresses at $T_g=0$, is unrelated to the critical state, which exists only in the presence of a finite granular agitation, $T_g=fv_s\not=0$. In Sec.~\ref{sapc}, we shall identify the angle of repose $\varphi_{re}$ with the critical angle. Again, the measured angle may vary because the stress is not uniform. 

In the space spanned by the density and stress components, the surface where the second derivative of the elastic energy changes its sign (turning from convex to concave) is a divide: On one side elastic solutions are stable, with grains at rest, on the other elastic solutions are instable, infinitesimal perturbations suffice to destroy any solution, and the grains are always agitated. 
The elastic energy of Eq~(\ref{2b-2}) is convex only for 
\begin{equation}\label{2b-3} u_s/\Delta\le\sqrt{2{\cal B}/{\cal A}} \quad
\text{or}\quad \pi_s/P_\Delta\le\sqrt{2{\cal A}/{\cal B}}.
\end{equation} 
It turns concave when this condition is violated. (The second condition may be derived by considering Eq~(\ref{2b-1}),  
showing $P_\Delta/\pi_s=\sqrt{{\cal B}/2{\cal A}}$ is minimal for
$u_s/\Delta=\sqrt{2{\cal B}/{\cal A}}$.) 

On a plane inclined by the angle $\varphi$,  with $y$ the depth of the granular layer on the plane and $x$ directed along the slope, we take the stress to be $\pi_{xx}=\pi_{yy}=\pi_{zz}=P_\Delta$, $\pi_{xy}=\pi_s/\sqrt2$, $\pi_{yz},\pi_{xz}=0$. Integrating $\nabla_j\pi_{ij}=g_i\rho$ assuming a variation only along $y$, we find $\pi_{xy}=g\sin\varphi\int\rho(y)dy$ and $\pi_{yy}= \pi_{xy}/\tan\varphi$. The angle of stability $\varphi_{st}$ (for infinite geometry) is reached when the energetic instability of Eqs.(\ref{2b-3}) is breached. With  $\pi_s^{yield}\equiv P\sqrt{2{\cal A}/{\cal B}}$ denoting the yield shear stress, it is
\begin{equation}\label{sb12}
\tan\varphi_{st}=\pi_s^{yield}/\sqrt2 P=\sqrt{{\cal A}/{\cal B}}.
\end{equation} 
Modifications from the proximity to walls or floor are not included here.

For the parameterization of the coefficients ${\cal A},{\cal B}$, we need a second  characteristic density, the {\it random-close} one, $\rho_{cp}$. It is the highest density at which grains may remain uncompressed. 
Assuming ${\cal B}/{\cal A}$ is density-independent (typically 5/3), and denoting $\bar\rho\equiv(20\rho_{\ell p}-11\rho_{cp})/9$, 
we take   
\begin{equation}\label{2b-4} {\cal B}={\cal B}_0
[(\rho-\bar\rho)/(\rho_{cp}-\rho)]^{0.15},
\end{equation}
with ${\cal B}_0>0$ a constant. This expression accounts for three  granular characteristics:
\begin{itemize}
\item The elastic energy is concave for $\rho<\rho_{\ell p}$, rendering all elastic solutions instable. 
\item The energy is convex between $\rho_{\ell p}$ and $\rho_{cp}$, ensuring the stability of elastic solutions in this region. In addition, the density dependence of sound
velocities (as measured by Harding and Richart~\cite{hardin}) is well
rendered by an expression  $\propto\sqrt{{\cal B}(\rho)}$. (The usual formula used to report the resuslts of~\cite{hardin} implies an unphysically  concave energy.) 
\item  The elastic energy diverges, slowly, for
$\rho\to\rho_{cp}$, approximating the observation that the system becomes quite a bit  stiffer there.
\end{itemize}

\subsection{Two Elastic Limits\label{elasticLimit}}
Because there is no space for granular rearrangements at the closed-packed density $\rho_{cp}$, plastic motions cannot take place and the system should be fully elastic.  Taking
$\alpha,\alpha_1,\lambda,\lambda_1\to0$ for $\rho\to\rho_{cp}$  in Eqs.(\ref{2c-7},\ref{2c-8},\ref{sigma}), reduces GSH to simple elasticity,
\begin{equation}\label{sigmaela}
\sigma_{ij}=\pi_{ij},\quad \partial_tu_{ij}=v_{ij}.
\end{equation} 
Close to $\rho_{cp}$,  the dependence on $\rho_{cp}-\rho$ is more sensitive than any  on $\rho$ directly. Hence we take, with $a_1,a_2,a_3,\lambda_0, \lambda_{10}>0$ and  $r\equiv1-\rho/\rho_{cp}$:
\begin{align}\label{DD}
\alpha=\alpha_0 r^{a_1},\quad \alpha_1=\alpha_{10}r^{a_2},\quad  \lambda/\lambda_0=\lambda_1/\lambda_{10}=r^{a_3}.
\end{align}
Granular media should also be fully elastic  for $T_g\to0$, 
see the  discussion at the end of the introduction, and in Sec.\ref{quasiEla}. 
In this  limit,  the stress increases steeply with the shear rate, until the yield stress is reached, at which point the system collapses. There is no stress relaxation, hence neither a critical state.

\subsection{More on $T_g$ and $u_{ij}$}
As $T_g$ and $u_{ij}$ are novel variables that entail conceptual subtleties, a discussion is useful for those who want to look beyond the present equations. It is also useful because Sun Qichen and coworkers~\cite{Sun} have variously proposed a ``Twin-Granular-Temperature Theory" that we find necessary to comment on. 

\subsubsection{The granular temperature $T_g$}
In a uniform medium (such as water or copper) there are two length scales, macro- and microscopic. All degrees of freedom may be 
divided into one of the two groups. A hydrodynamic theory  takes the degrees of freedom from 
the first group as explicit variables, including mass, momentum and energy density, also the strain field in the case of solids. 
All degrees of freedom from the second group are taken summarily, with their contribution to the energy lumped together as 
heat, and characterized by the temperature $T$.

In granular media, there is an intermediate group, consisting of momentum and deformation of each grain. In the kinetic theory or DEM, these are taken as independent variables. But a summary inclusion again suffices here, with the associated energy lumped into granular heat, characterized by $T_g$. 
As the kinetic energy changes rapidly into an elastic one, and back, during a collision, both types of energy interact, equilibrate, and must be lumped into one single macroscopic variable $T_g$. 
On the other hand, 
distinguishing $T_g$ from $T$ is necessary: $T_g$ is  many orders of magnitude higher than $T$ in elasto-plastic motion, and when this is the case, granular stress relaxes, giving rise, as discussed, to plasticity. 

There are also grain-sized deformations, such as in force chains, that are static or slowly varying in time. Their contribution is not included in $T_g$, but in that of  $u_{ij}$, see below.

\subsubsection{The elastic strain $u_{ij}$}

Accounting for the macroscopic elastic stress that varies slowly in space, the elastic strain is also slowly varying. As energy is positive and conserved, the macroscopic energy density $w$ is the sum of all (slowly varying) mesoscopic contributions, $w^{mes}$. In other words,  $w$ is the coarse-grained mesoscopic one, $w=\langle w^{mes}\rangle$, and since the latter includes the contribution from the force chains, the former also does. 

It is important to realize that  the elastic strain cannot be obtained by coarse-graining, $u_{ij}\not=\langle u_{ij}^{mes}\rangle$. This is because both the energy and the elastic stress are coarse-grained quantities:  $w=\langle w^{mes}\rangle$ and  $\pi_{ij}=\langle\pi_{ij}^{mes}\rangle$. With
${\rm d}w=\langle{\rm d}w^{mes}\rangle=-\langle \pi_{ij}^{mes}{\rm d}u_{ij}^{mes}\rangle=-\langle \pi_{ij}^{mes}\rangle{\rm d}u_{ij}$ and $\langle \pi_{ij}^{mes}{\rm d}u_{ij}^{mes}\rangle\not=\langle \pi_{ij}^{mes}\rangle{\rm d}\langle u_{ij}^{mes}\rangle$, we conclude: $u_{ij}\not=\langle u_{ij}^{mes}\rangle$.

\subsubsection{Rebuttal of ``Twin-Granular-Temperature Theory"\label{tgt}}
Sun and coworkers have written a number of papers~\cite{Sun} proposing a generalization of GSH by introducing two granular temperatures, $T^k$ and $T^c$, with  $T^k$ taking account of the kinetic, and  $T^c$ the elastic energy (including contributions from force chains). 
It should be clear from the last two sections that neither is a well-behaved macroscopic variable. The temporally fast varying elastic energy from collisional deformation is necessarily part of $T_g$, while the slowly varying contribution from force chains are part of $u_{ij}$.

\subsection{Summary\label{summary}}

  With the differential equations given and the energy density specified, GSH is a fairly well-defined theory. It contains clear ramifications and provides little leeway for retrospective adaptation to observations.  
As we shall see in the rest of the paper, a wide range of granular phenomena is well encoded in these equations.

The three relaxation equations, (\ref{Tg2},\ref{eqU},\ref{eqD}), possess a steady state solution: 
$T_g=fv_s$, $u_s=u_c$, $\Delta=\Delta_c$, which is simply the critical state. With it given, we may consider the relaxation dynamics -- either at  given shear rate, or holding the shear stress constant. In the first case, 
because of the small relaxation time $1/R_TT_g$, we have $T_g=fv_s$ practically instantly, rendering $T_g$ dependent. Then the strain relaxation equation is manifestly rate-independent,  displaying the mathematical structure of  hypoplasticity. 
The rich phenomenology observed in triaxial apparatus is then well accounted for by hypoplasticity and GSH alike.
The relaxation dynamics at constant shear stresses gives rise to phenomena such as creeping, shear band, angle of repose --  all beyond hypoplasticity. They are considered in Sec.\ref{aging}.

\section{Experiments at Given Shear Rates\label{hypoplastic motion}}

\subsection{The Critical State\label{critical state}} 
The physics of the critical state is: Deformed grains give rise to an  elastic stress. 
When grains are sheared past one another, they  rattle and jiggle. And when they do, 
the deformation and stress is slowly lost, because of the loosen contact to their neighbors.  As a 
consequence, a shear rate not only increases
the deformation, as in any elastic medium, but also
decreases it.  A steady state exists in which both
processes balance, such that the elastic deformation and stress remain
constant over time. This is the critical state. Moreover, the  critical stress is rate-independent, because (1)~the increase in deformation is $\propto v_s$, (2)~the  relaxation is $\propto T_g$, and (3)~$T_g\propto v_s$ in $T_g$'s steady state. 

\subsubsection{A Stationary Elastic Solution\label{Stationary Elastic Solution}}
Taking $\partial_t T_g,\, \partial_t u_{ij}^*,\,\partial_t\Delta=0$ in Eqs.(\ref{Tg2},\ref{eqU},\ref{eqD}) for $v_{ij}^*=$ {const}, $v_{\ell\ell}=0$, we have
\begin{align} T_g&=fv_s,\quad  \Delta=\Delta_c.\label{cs1}
\\\label{cs2}  u_{ij}^*&=u_c\hat v_{ij},\quad \text{or}\,\,\, u_s=u_c, \,\, \hat u_{ij}=\hat v_{ij}.
\end{align}
With $\Delta=\Delta_c$, $u_{ij}^*=u_c\hat v_{ij}$ and $\rho$ being constant, so is the associated elastic stress $\pi_{ij}(u_{k\ell},\rho)$. 
The stress value is obtained by inserting Eqs.(\ref{cs1},\ref{cs2}) into Eqs.(\ref{2b-2a},\ref{2b-2b},\ref{2b-1}),
\begin{align}
\label{3b-4a}
P_c=(1-\alpha)P^c_\Delta,\quad
\sigma_c=(1-\alpha)\pi_c, 
\\\label{3b-3c}
P^c_\Delta\equiv P_\Delta(\Delta_c,
u_c),\,\, 
\pi_c\equiv\pi_s(\Delta_c, u_c), \\\label{3b-3cA}
{P^c_\Delta}/{\pi_c}=({{\cal B}}/{2{\cal
A}}){\Delta_c}/{u_c}+{u_c}/{4\Delta_c}. 
\end{align} 
The loci of the critical states thus calculated~\cite{GSH&Barodesy}  (though employing the slightly more general energy of~\cite{3inv}) greatly resembles those calculated using either hypoplasticity or barodesy~\cite{barodesy}

Same as the Coulomb yield of Eq~(\ref{sb12}), the critical ratio $\sigma_c/P_c=\pi_c/P_\Delta$ is also frequently associated with a friction angle. Since the former is relevant for vanishing $T_g$, while the latter requires an moderately elevated $T_g\propto v_s$, it is appropriate to identify one as the static friction angle, or yield angle, and the other as the
dynamic one, or the critical angle. In Sec.~\ref{sapc}, we shall present reasons why the critical angle should be identified with the angle of repose $\varphi_{re}$, the angle at which granular flows will come to a halt for uniform stress.

Textbooks on soil mechanics frequently state that the friction angle is
 independent of the density -- although they do not usually 
distinguish between the dynamic and the static one. We shall assume, for lack of better  information, that both are.

\subsubsection{Approach to the Critical State at Constant Density\label{approach critical}} 
Solving  Eqs.(\ref{Tg2},\ref{eqU},\ref{eqD}) for $u_s(t),\Delta(t)$, at constant
$\rho, v_s$, with the initial conditions: $\Delta=\Delta_0, u_s=0$ (same as $P=P_0,\sigma_s=0$), the relaxation into the critical state is given as
\begin{align}\label{3b-6}
u_s(t)=u_c(1-e^{-\lambda f\varepsilon_s}),\quad \varepsilon_s\equiv v_st,
\\\nonumber \Delta(t)=\Delta_c(1+f_1\,e^{-\lambda f
\varepsilon_s}+f_2e^{-\lambda_1f\varepsilon_s}), \\\nonumber
f_1\equiv\frac{\lambda_1}{\lambda-\lambda_1},\quad
f_2\equiv\frac{\Delta_0}{\Delta_c}-\frac{\lambda}{\lambda-\lambda_1}.\end{align}
This is an exponential decay for $u_s$, and a sum of two decays for $\Delta$. It is useful, and quite demystifying, that a simple, analytical solution exists. 

Typically, we take $\lambda/\lambda_1\approx3$, as the decay of $u_s$ and $f_1$ are 
faster than that of $f_2$. The prefactor $f_2$ may be
negative, and $\Delta(t)$ is then non-monotonic. The associated pressure
and shear stress are those of Eqs~(\ref{3b-4a},\ref{3b-3c},\ref{3b-3cA}). For a negative $f_2$, neither the pressure nor the shear stress is monotonic. 
For the system to complete the approach to the critical state, the yield surface of Eq.(\ref{2b-3}) must not be breached during the non-monotonic path.

\subsubsection{Approach to the Critical State at Constant Pressure\label{pressure approach}}
\begin{figure}[t]
\includegraphics[scale=.9]{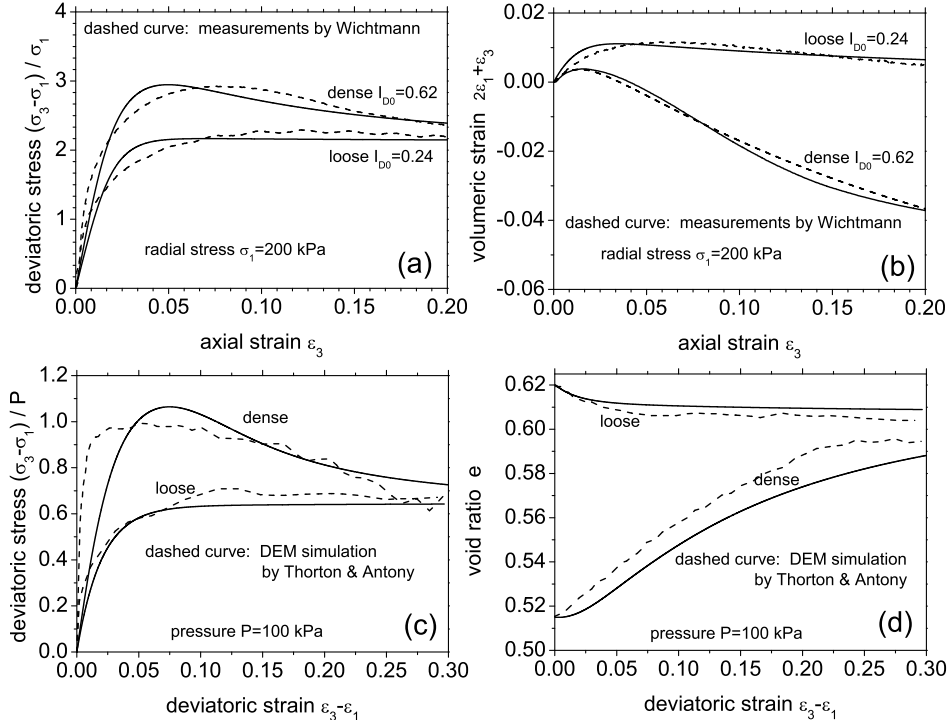}
\caption{A GSH  calculation employing the parameters of I for comparing to the Wichtmann's~\cite{wichtmann} (drained monotonic triaxial)  experiment, and II to the simulation by Thornton and Antony~\cite{thornton}, both in the plots as originally given. } \label{fig3a}
\end{figure}
Frequently, the critical state is approached at constant pressure $P$, or a stress eigenvalue $\sigma_i$. The circumstances are then more complicated. As $\Delta,u_s$ approach
$\Delta_c,u_c$, the density compensates to keep
$P(\rho,\Delta,u_s)=$ const. Along with $\rho$, the coefficients
$\alpha,\alpha_1,\lambda,\lambda_1,f, \Delta_c,u_s$ (all functions of $\rho$), also change
with time. In addition, with $\rho$ changing, the compressional flow
$v_{\ell\ell}=-\partial_t\rho/\rho$ no longer vanishes (though it is still small). Analytic solutions
do not seem feasible, but numerical ones are, see Fig \ref{fig3a}. The parameters, labeled as I and  II, are:
\begin{itemize}
\item ${\cal B}_0=2, 0.22$ GPa,\quad  ${\cal B/A}=5/3, 8$,
\quad $\bar\rho/\rho_{cp}=0.615, 0.650$, 
\item $\alpha_0=1.04, 0.85$,\quad $\alpha_{10}=400, 30$,\quad $\lambda_0f=272, 250$,\quad $\lambda/\lambda_1=3.8, 3.8$, 
\item $a_1=0.15, 0.15$,\,\, $a_2=1,0. 15$,\,\, $a_3=0.6, 0.53$.
\end{itemize}

\subsubsection{Perturbations and Rate-Dependence of the Critical Stress \label{external perturbation}}
\begin{figure}[t]
\includegraphics[scale=.8]{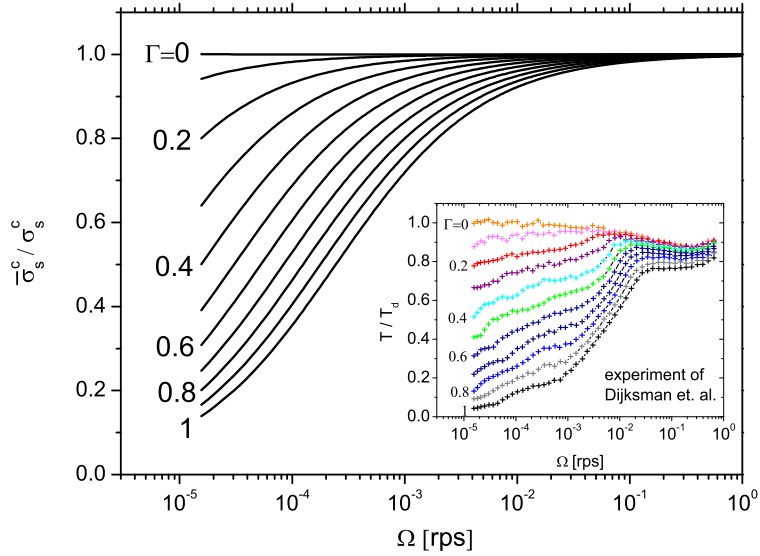}
\caption{Suppression of the critical shear stress $\sigma_c$ by vibration as given by Eq.(\ref{3b-8}) [assuming $\Gamma=\alpha T_a,\,\, \Omega=\beta v_s^3$, see~\cite{granRexp} for details]. Inset is the experimental curve of~\cite{vHecke2011}, with the torque $\tau$ denoted as $T$, as in~\cite{vHecke2011}.  (The stress dip at large $\Omega$, neglected here, is explained in~\cite{StressDip}.) }\label{fig3b}
\end{figure}

If one perturbs the system, say by exposing it to weak vibrations,  or by tapping it periodically, such as in a recent experiment~\cite{vHecke2011}, the critical state is modified, and a strong rate-dependence of the critical shear stress is observed. The stress decreases with the tapping amplitude, and increases with the shear rate, such that the decrease is compensated at high rates. Clearly, any theory with built-in rate-independence cannot account for this observation. In {GSH}, on the other  hand, rate-independence is a result of $fv_s/T_g=1$, and expected to be destroyed by any ambient temperature $T_a\ne0$, see Eq.(\ref{fVs}). The steady state values are then reduced to $\bar u_c\equiv(fv_s/T_g)u_c$,  $\bar \Delta_c\equiv(fv_s/T_g)^2\Delta_c$,   $\bar\sigma_c/\sigma_s= {\bar\Delta_c^{0.5}}\bar u_c/{\Delta_c^{0.5}}u_c=(fv_s/T_g)^2$,  see Eqs.(\ref{eqU},\ref{eqD}), implying a rate-dependence via Eq.(\ref{fVs}),
\begin{equation}\label{3b-8}\frac{\bar u_c^2}{u_c^2}=
{\frac{\bar \Delta_c}{\Delta_c}}={\frac{\bar \sigma_c}{\sigma_c}}=\frac1{1+{ T_a^2}/{(fv_s)^2}}.
\end{equation}
If there is no tapping, $T_a=0$, we retrieve the unperturbed values, $\bar u_c=u_c$, $\bar\Delta_c=\Delta_c$, $\bar\sigma_c=\sigma_c$. With tapping,  $\bar u_c, \bar\Delta_c, \bar\sigma_c$ decrease for increasing $T_a$, and increase with increasing shear  rate $fv_s\equiv f|v_s|$. see Fig~\ref{fig3b}. 
(Note we have only considered the critical state at given shear rate, not the approach to it. So the result holds both at given density and pressure.)

The above consideration is the basic physics of the observation of~\cite{vHecke2011}. It helps to put rate-independence, frequently deemed a fundamental property of granular media, into the proper context. 
A quantitative comparison is unfortunately made difficult by the highly nonuniform experimental geometry. 
Nevertheless, some comparison, even if unabashedly qualitative, may still be useful. 
In~\cite{vHecke2011}, the torque $\tau$ on the disk on top of a split-bottom shear cell is observed as a function of  its rotation velocity $\Omega$ and the shaking acceleration $\Gamma$. Now, $(\tau, \sigma_s)$,  $(\Omega, v_s)$, $(\Gamma, T_a)$ are clearly related pairs.
Assuming the lowest order terms suffice in an expansion, we take $\sigma_c\propto\tau$, $\Gamma\propto T_a$.   If $v_s$ were uniform, $\Omega\propto v_s$ would also hold. Since it is not, $\Omega\propto v_s^n$, $n>1$ seems plausible, because with additional degrees of freedom such as position and width of the shear band, the system has (for given $\Omega$) more possibilities to decrease its strain rate $v_s$. We take $\Omega=c_2v_s^3$ with $c_2=1{\rm rs}^2$ [replacing $T_a/v_s$ with $\Gamma/\sqrt[3]\Omega$ in Eq.(\ref{3b-8})]   to fit Fig \ref{fig3b}, emphasizing that qualitative agreement exists irrespective of the fit.  

\subsubsection{Load and Unload\label{Load and Unload}} 
Inserting $T_g= fv_s$, $v_{\ell\ell}=0$ into Eqs~(\ref{eqU},\ref{eqD}), keeping the shear rate along one direction  and allowing reversal, $\eta\equiv\hat v_{ij}\hat u_{ij}=\pm1$, we have
\begin{align}\label{3b-2}
\partial_t\Delta/v_s=-\lambda_1 f(\Delta-\eta \Delta_cu_s/u_c),
\\\label{3b-3}
\partial_tu_s/v_s=-\lambda f(u_s-\eta u_c),
\end{align} 
and conclude that the reason for the difference in apparent stiffness: $\partial_tu_s/v_s$ and $\partial_t\Delta/v_s$, between load and unload, is that $\eta$ turns negative when the shear rate $v_{ij}^*$ is reversed. The strain then relax towards the new stationary values, with a simple relaxation dynamics not at all {\it history-dependent}. 
That loading ($\eta=1$) and unloading ($\eta=-1$) have different slopes is frequently referred to as {\em
incremental nonlinearity} in soil mechanics. It is the reason why no backtracing takes place under reversal of shear rate. The stress components $P,\sigma_s$ are calculated employing Eqs~(\ref{2b-2b}) for given $\Delta, u_s$ and $\rho$.
\begin{figure}[t] \begin{center}
\includegraphics[scale=0.35]{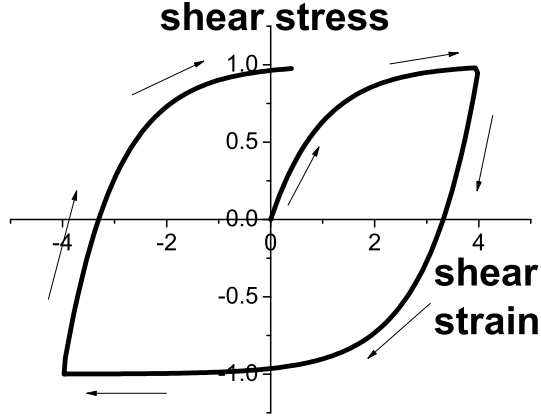}
\end{center}
\caption{\label{fig2}The change of the elastic shear strain $u_{s}/u_c$ with the total strain $\varepsilon\lambda f$, as given by Eq~(\ref{3b-3}).} 
\end{figure} 

\subsection{Hypoplasticity,  Neo-Hypoplasticity and Barodesy\label{constRel}}

As discussed around Eqs.(\ref{dot u},\ref{dot u2}), GSK possesses the hypoplastic structure for given shear rates, $T_g=fv_s$. So, this is a lucky instance in which a comparison may be made directly, by comparing the coefficients -- where one set is proposed from accumulated knowledge of granular behavior, while the other is given by the second derivative of the elastic energy. This is done in Sec~\ref{Hypoplasticity}. Then we show that all recent hyperelastic models, either given as an elastic energy or a Legendre potential, are rather similar if properly transformed. The main difference in fact stems from whether or not there are elastic instabilities.  

Barodesy, a recent model again proposed by Kolymbas~\cite{barodesy}, has a rather more complicated rate dependence than the one shared by hypoplasticity and GSH. But our comparison of  GSH and barodesy did yield essentially quantitative agreement~\cite{GSH&Barodesy,P&G2009}. 
The idea of barodesy is to have a more modular and better organized structure than hypoplasticity, with different parts in the constitutive relation taking care of specific aspects of granular deformation, especially that of {\em proportional paths} as summed up by  the Goldscheider rule. In Sec.~\ref{pep}, we shall focus on the latter showing how  the Goldscheider rule and proportional paths naturally arise from GSH.

\subsubsection{A Comparison of the Hypoplastic Coefficients\label{Hypoplasticity}}

Writing the hypoplastic model as
\begin{eqnarray}
\partial _{t}\sigma _{ij}^{\ast } &=&H_{1}\sigma _{nn}v_{ij}^{\ast }+H_{2}%
{\left( \sigma _{lk}^{\ast }v_{lk}^{\ast }\right) \sigma _{ij}^{\ast }}/{\sigma _{nn}}
+H_{5}v_{nn}\sigma _{ij}^{\ast }-H_{6}\sigma _{ij}^{\ast }\sqrt{%
v_{lk}v_{lk}}  \label{HYP1a} \\\nonumber
\partial _{t}\sigma _{nn} &=&H_{3}v_{nn}\sigma _{mm}+H_{4}\left( \sigma
_{lk}^{\ast }v_{lk}^{\ast }\right) -H_{7}\sigma _{nn}\sqrt{v_{lk}v_{lk}}  
\end{eqnarray}
we have the dimensionless coefficients $H_{1-7}$ that are functions of $\rho$ 
and the 3 stress invariants: $P\equiv \sigma _{nn}/3$, $\sigma _{s}\equiv\sqrt{\sigma _{ij}^*\sigma _{ij}^*}$, $\sigma _{t}\equiv\sqrt[3]{\sigma _{ij}^*\sigma _{jk}^*\sigma _{ki}^*}$. In~\cite{Nie3}, the authors present the parameter values for more than forty different granular media, with 
\begin{eqnarray}\label{2}
H_{1} &=&-f_{0}F^{2},\quad  
H_{2} =-f_{0}a^{2}, \quad
H_{3} =-f_{0}F^{2}-{f_{0}a^{2}}/{3}, \quad H_{4}=-f_{0}a^{2}, 
\\
H_{5} &=&-{f_{0}a^{2}}/{3}, \quad
H_{6} =2f_{0}af_{d}F, \quad
H_{7} =f_{0}af_{d}F, 
\end{eqnarray}
characterized by three functions $F,f_0,f_d$ and one constant $a$, given as 
\begin{align}
f_{0}&=\frac{\sigma _{nn}f_{b}f_{c}}{\sigma _{lk}\sigma _{lk}},\quad
f_{d}=\left( \frac{e-e_{d}}{e_{c}-e_{d}}\right) ^{\alpha },\quad
a =\sqrt{\frac{3}{8}}\,\frac{3-\sin \varphi _{c}}{\sin \varphi_{c}}, 
\\
F &=\sqrt{\frac{\tan^2 \psi}{8}+\frac{2-\tan^2\psi}{2+\sqrt{2}\tan \psi \cos 3\theta }}-\frac{\tan
\psi }{2\sqrt{2}}, \nonumber
\end{align}
where
\begin{align*}
f_{c}&=\left( \frac{e_{c}}{e}\right) ^{\beta }, \quad 
\cos 3\theta =\sqrt{6}\left( \frac{\sigma _{t}}{\sigma _{s}}\right) ^{3}, \quad
\tan\psi =\frac{\sigma _{s}}{\sqrt{3}P}, 
\\
f_{b}&=\left[\frac{e_{i0}}{e_{c0}}\right]^{\beta }\frac{h_{s}}{n}  
\frac{e_{i}+1}{e_{i}} \left[\frac{3P}{h_{s}}\right] ^{1-n}
\left[ 3+a^{2}-\sqrt{3}a\left( \frac{e_{i0}-e_{d0}}{e_{c0}-e_{d0}}\right)
^{\alpha }\right] ^{-1}, \\
e_{d}&=e_{d0}f_{e}, \quad e_{c}=e_{c0}f_{e},\quad
e_{i}=e_{i0}f_{e}, \quad f_{e}=\exp \left[ -\left( \frac{3P}{h_{s}}\right) ^{n}\right], 
\end{align*}
contain eight material constants, with $\varphi_c$ the critical frictional angle,  and $e_{0d},e_{0c},e_{0i}$ denoting the densest, critical and loosest void ratio. For the {\em Stuttgart sand}, these are  given as 
\begin{align}\label{table}
{\varphi _{c}=0.576,\,\, h_{s}=2600\text{MPa},\,\, n=0.3,\,\, \alpha =0.1,\,\, \beta =1,}
\\\nonumber
{e_{0d}=0.6,\,\, e_{0c}=0.98,\,\, e_{0i}=1.15.}\qquad\qquad
\end{align}
Note $\sigma _{ij}$ is here negative from that in~\cite{Nie3}; 
note also that the stress invariant $\sigma _{t}$ in $\cos 3\theta$ and  $F$
will be approximated below with $\cos 3\theta =1$, ie.  $\sigma _{t}^{3}=\sigma _{s}^{3}/\sqrt{6}$, valid for 
axial compressions.

Returning to GSH, see Eqs.(\ref{eqU},\ref{eqD}) with $T_g=fv_s$, we have four dynamic coefficients,
\begin{equation}
\alpha,\,\,\Lambda\equiv\lambda f,\,\,\epsilon\equiv\lambda_1/\lambda,\,\,\alpha_1\,\, \text{or}\,\, \Delta_c/u_c=\alpha_1/\lambda_1 f,
\end{equation}
and two static coefficient $\cal B$ and ${\cal B/A}=5/3$. 

Defining an operator that is valid for any function $f(\Delta,u_s,\rho)$ such as the stress $\sigma_{ij}$, 
\begin{align}
\partial _{t}&=V\partial _{1}+\frac{\sigma _{ij}v_{ij}^{\ast }}{P}\partial
_{2}+v_{s}\partial _{3}, \quad\text{where} \label{19}
\\\nonumber 
\partial _{1} &\equiv\rho \frac{\partial }{\partial \rho }+\left( 1-\alpha
\right) \frac{\partial }{\partial \Delta},   \\
\partial _{2} &\equiv\frac{\alpha _{1}P}{\sigma _{1}}\frac{\partial }{\partial
\Delta }+\frac{\left( 1-\alpha \right) P}{\sigma _{1}u_{s}}\frac{\partial }{%
\partial u_{s}},  \nonumber \\
\partial _{3} &\equiv-\Lambda u_{s}\frac{\partial }{\partial u_{s}}-\epsilon
\Lambda \Delta \frac{\partial }{\partial \Delta },  \nonumber
\\\nonumber
\sigma&\equiv -\left( 1-\alpha \right) 2A\sqrt{\Delta }-\alpha
_{1}P_{\Delta },\quad \sigma _{ij}^{\ast }=\sigma u_{ij}^{\ast },
\end{align}%
we rewrite
\begin{align*}
\partial _{t}\sigma _{ij} &=V\left( \partial _{1}P\right) \delta
_{ij}+\left( \sigma _{lk}v_{lk}^{\ast }\right) \left( \partial _{2}P\right)
\delta _{ij}+v_{s}\left( \partial _{3}P\right) \delta _{ij}   \\
&+V\frac{\partial _{1}\sigma }{\sigma }\sigma _{ij}^{\ast }+\left(
\sigma _{lk}v_{lk}^{\ast }\right) \frac{\partial _{2}\sigma }{\sigma}
\sigma _{ij}^{\ast }+v_{s}\frac{\partial _{3}\sigma}{\sigma}%
\sigma _{ij}^{\ast } \\
&+\sigma\left( 1-\alpha \right) v_{ij}^{\ast }-\lambda T_{g}\sigma
_{ij}^{\ast },
\end{align*}
 in the form of Eq.(\ref{HYP1a}) to obtain
\begin{align}
H_{1} &=\frac{\sigma\left( 1-\alpha \right) }{3P},  \quad
H_{2} =\frac{\partial _{2}\sigma}{\sigma},  \quad
H_{3} =-\frac{\partial _{1}P}{P},  \quad
H_{4} =3\frac{\partial _{2}P}{P}, \\
H_{5} &=-\frac{\partial _{1}\sigma}{\sigma},  \quad
H_{6} =\Lambda -\frac{\partial _{3}\sigma}{\sigma},  \quad
H_{7} =-\frac{\partial _{3}P}{P}. 
\end{align}
We first parameterize the static coefficients. Using the expression for the energy, Eqs.(\ref{2b-2},\ref{2b-4}), we take ${{\cal A/B}=5/3}$ and ${{\cal B}_0=0.4h_s}$, with $h_s$, $e_{0i}=\rho_G/\rho_{lp}-1$, $e_{0d}=\rho_G/\rho_{cp}-1$ given by table~\ref{table}. (The bulk density $\rho_G$ does not enter the formula.) Next, we take the dynamic coefficients to be  ${\lambda f=600r^{0.5}}$, where $r\equiv(\rho_{cp}-\rho)/\rho_{cp}$; ${\lambda_1 f=0.3\cdot\lambda f}$; and 
${\alpha=(\rho_{lp}/\rho)^2(1-\rho_{lp}/\rho_{cp})^{-2}\, (1-r)\, r^2}$.  Finally, we take ${\alpha_1=0.103 \lambda f}$ (because with $\alpha_1/\lambda_1 f=\Delta_c/u_c$, we obtain $\pi_c/P_\Delta^c=1,67\tan\varphi_c$, in agreement with $\varphi$ as given in table~\ref{table}.) We note that the time scale $R_T$ and length scale $\xi_T$ of Eq.(\ref{Tg2}) are not contained in the hypoplastic model, and their values therefore not needed for the comparison. 
\begin{figure}[t] \begin{center}
\includegraphics[scale=.85]{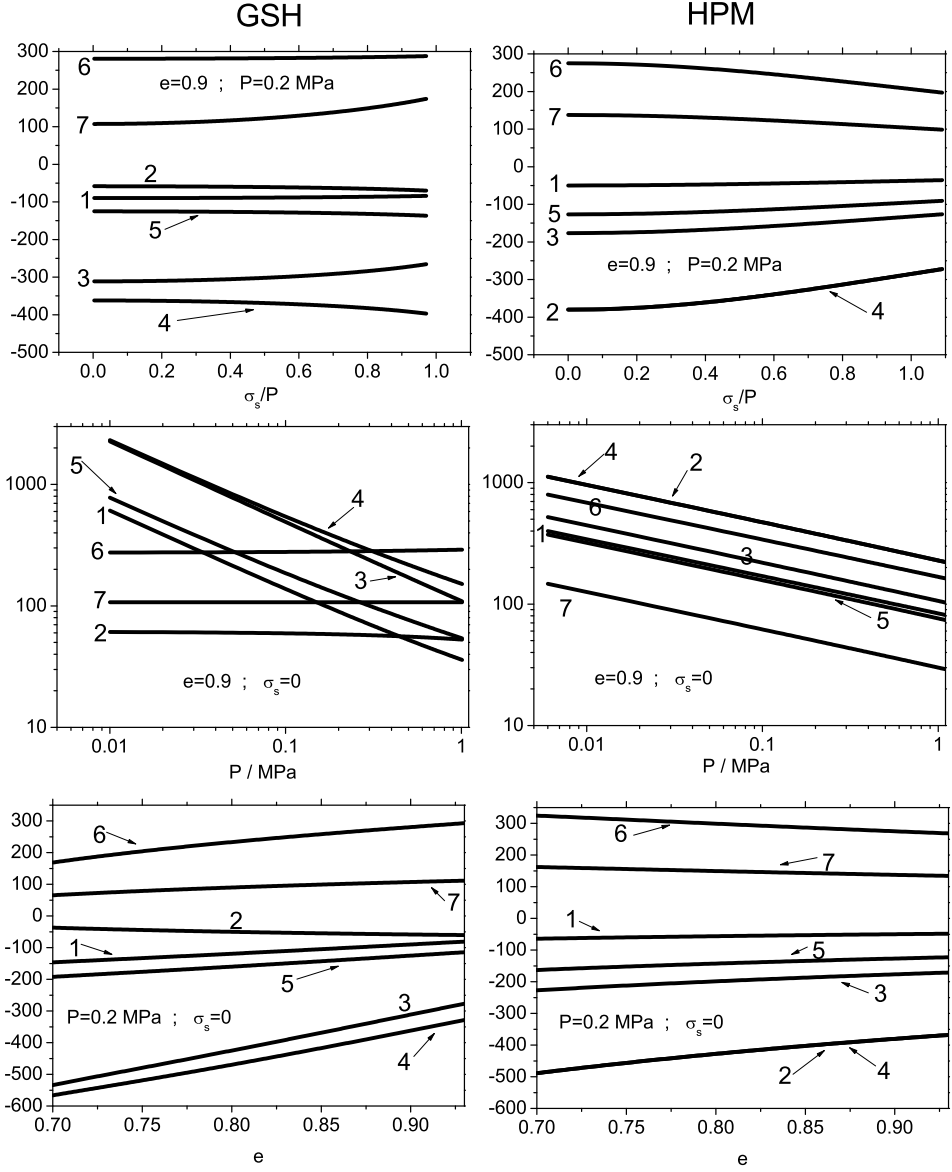}
\end{center}
\caption{\label{hypopla} Comparison of the coefficients $H_{1-7}$ between GSH
and HPM:  (1) as functions of $\sigma_s/P$, at given $e$ and $P$; (2) as functions of $P$, at given $e$ and $\sigma_s=0$; (3) as functions of $e$, at given $P$  and $\sigma_s=0$. Note $H_{2}=H_{4}$ for HPM. For comments see text.} 
\end{figure} 

Although qualitative agreement is obvious from Fig~\ref{hypopla}, a quantitative one is lacking. We do not take this as a serious defect of GSH, for three reasons. First, complete agreement of coefficients is probably an unnecessarily difficult task: Different coefficients frequently yield similar experimental curves. 
Second and more importantly, the parameterization of GSH, especially its energy, has been chosen stressing manipulative ease and simplicity of expressions. GSH's robust and qualitative result is the fact that hypoplastic coefficients may be obtained from the second derivative of a material-dependent scalar potential. 

Third, starting with the Gibbs potential and choosing $n=0.6,\alpha=0.1$,
\begin{equation}\label{einav}
G\propto{P_\Delta^{2-n}}\left[1+\frac13 \frac{\pi_s^2}{P_\Delta^2}\right]^{1-\frac{n+\alpha}2}, 
\quad u_{ij}=\frac{\partial G}{\partial\pi_{ij}},
\end{equation}
Niemunis, Grandas Tavera and Wichtmann most recently use its second derivative to obtain the coefficient $H_{ijkl}$ of Eq.(\ref{3b-1}) -- a model that they call {\em neo-hypoplasticity}~\cite{99}. This is clearly a step that we highly welcome.  Calibrating the coefficients using incremental stress-strain relations, they found good agreement with observation. 
The next three sections are about comparing four more recent hyperelastic models, to help optimize the choice of the energy.

\subsubsection{Elastic Energy versus the  Gibbs Potential}
Elatic constitutive relations that possess an explicit elastic energy or Gibbs potential is usually termed a hyperelastic theory. Engineers tend to look for appropriate Gibbs potentials, because they prefer the stress as variable. On the other hand, only the energy is conserved and useful for the hydrodynamic procedure. Although one can obtain one from the other, simple energy expressions typically possess complicated Gibbs potentials, and vice versa.  In this section, we collect some general results connecting both, starting with the basic one of the Legendre transformation:%
\begin{equation}
\psi =w+u_{ij}\pi _{ij},\quad u_{ij}={\partial \psi }/{\partial \pi _{ij}},\quad \pi _{ij}=-{\partial w}/{\partial u_{ij}}.  \label{psi-w}
\end{equation}
Assuming the elastic energy is a homogeneous function of degree $m$, we have \begin{equation}
w\equiv B\Delta^{m}{\cal F}\left(\widehat{u}_{s}\right), \quad \widehat{u}_{s}\equiv {u}_{s}/\Delta,
\label{w=F}
\end{equation}%
with $B$ a constant. This 
leads immediately to 
\begin{eqnarray}
\pi _{ij} &=&B\Delta ^{m-1}\left( m{\cal F} -\widehat{u}_{s}{\cal F}
^{\prime }\right) \delta _{ij}-B\Delta ^{m-1}{\cal F} ^{\prime }(u_{ij}^{\ast }/u_{s}),  \label{pi-ij-homogenous} \\
P_{\Delta } &=&B\Delta ^{m-1}\left( m{\cal F} -\widehat{u}_{s}{\cal F}
^{\prime }\right),   \label{pi-homogenous} \quad
\pi _{ij}^{\ast } =-B\Delta ^{m-1}{\cal F} ^{\prime }(u_{ij}^{\ast }/u_{s}),  \label{pi-ij-star-homogenous} \\
\pi _{s} &=&B\Delta ^{m-1}{\cal F} ^{\prime },  \label{pi-s-homogenous} 
\qquad
\widehat{\pi }_{s}\equiv \pi _{s}/P_{\Delta }={{\cal F} ^{\prime }}/({m{\cal F} -\widehat{u}_{s}{\cal F} ^{\prime }})
\label{pi-ratio-homogenous}
\end{eqnarray}%
(where prime $^{\prime }$ means derivative).
The Legendre transformed Gibbs potential is then
\begin{equation}
\psi =\left( 1-m\right) B\Delta ^{m}{\cal F}(\widehat{u}_{s}),   \label{psi-homogenuous}
\end{equation}%
with $\widehat{u}_{s},\Delta$  taken as functions of $P_{\Delta },\widehat{\pi }_{s}$, especially
$\Delta =\left( BP_{\Delta }\,{{\cal F}
^{\prime }}/{\widehat{\pi }_{s}}\right) ^{\frac{1}{1-m}}.$  

A remarkble property of the above energy expression is the simple relation,
\begin{equation}
{w}=(1-m){\psi }.  \label{w/psi}
\end{equation}%
As a result, the Gibbs potential  also has the factorized form:%
\begin{equation}
w=B^{\frac{1}{1-m}}(P_{\Delta })^{\frac{^{m}}{m-1}}F,   \label{w=w(pi)}
\end{equation}%
where the dimensionless shearing factor $F =F \left( \widehat{u}
_{s}(\widehat\pi_s)\right) $ is
\begin{equation}
F =\left( {\cal F} ^{\prime }/\widehat{\pi }_{s}
\right) ^{\frac{m}{1-m}}{\cal F}   \label{Im=F}.
\end{equation}%
For the Hertz contact $m=5/2$, 
we have $w=B^{-2/3}P_{\Delta }^{5/3}F $.

\subsubsection{Comparing Four Hyperelastic Models\label{neo-hypo}}
We compare four hyperelastic models, 
\begin{equation}
\begin{array}{cc}
\text{GSH} & w\sim \Delta ^{2+a}\left( 1+\widehat{u}_{s}^{2}/{\xi }%
\right),  \\ 
\text{Einav-Puzrin} & \psi \sim P_{\Delta }^{3/2}\left( 1+\widehat{\pi }%
_{s}^{2}/b\right),  \\ 
\text{Houlsby-Amorosi-Rojas} & w\sim \Delta ^{3}\left( 1+b\widehat{u}%
_{s}^{2}\right) ^{3/2}, \\ 
\text{neo-hypoplasticity} & \psi \sim P_{\Delta }^{2-n}\left( 1+\widehat{\pi }%
_{s}^{2}/3\right) ^{1-\frac{n+\alpha }{2}},
\end{array}
\label{GSH-other-elasticity}
\end{equation}%
with ${\xi }=2{\cal B}/5{\cal A}$ a density independent constant. 
GSH's  elastic energy (with $a$ unspecified) was proposed and considered in 2003, see~\cite{granL1}. In our later works, we concentrated on  $a=1/2$, the Hertz contact value, to be definite, and because Hertz contact seems the appropriate granular picture. 
Three works on hyperelastic models appeared since, all of the form  Eq.(\ref{w=F}), by Houlsby, Amorosi and Rojas~\cite{HousbyA},  Einav and Puzrin~\cite{EinavP}, and, as already mentioned, as neo-hypoplasticity, by Niemunis et al~\cite{99}.
We first note that the power of the ``spheric
part'' $P_{\Delta }^{\frac{^{m}}{m-1}}$ of Eq.(\ref{w=w(pi)})) are 1.5 for both Einav-Puzrin and Houlsby-Amorosi-Rojas,
1.4 for neo-hypoplasticity, and  $\frac{2+a}{1+a}$ for GSH, ie. 5/3 for a=0.5, and again 1.5 for a=1.

Calculating the  stress-strain relations for the two energies and their inverse, we have
\begin{equation}
\begin{array}{cc}
\text{GSH} & \widehat{\pi }_{s}=2\widehat{u}_{s}\left( (2+a)
{\xi }+a\widehat{u}_{s}^{2}\right) ^{-1}, \\ 
\text{Houlsby-Amorosi-Rojas} & \widehat{\pi }_{s}=b\widehat{u}_{s},%
\\
\text{GSH} & \widehat{u}_{s}=\left( 2+a\right) {\xi }\theta 
\widehat{\pi }_{s}/2, \\ 
\text{Houlsby-Amorosi-Rojas} & \widehat{u}_{s}=\widehat{\pi }_{s}/b.%
\label{GSH-HAR-u=pi}
\end{array}
\end{equation}%
The shearing factor $F $ for these models are therefore, see Eq.(\ref{Im=F}),
\begin{equation}
\begin{array}{cc}
\text{GSH} & F /F _{0}=\left[ 1+\left( 2+a\right) ^{2}{\xi }%
\widehat{\pi }_{s}^{2}\theta ^{2}/4\right] \theta ^{\frac{2+a}{-1-a}}, \\ 
\text{Einav-Puzrin} & F /F _{0}=\left( 1+\widehat{\pi }_{s}^{2}/b\right),
\\ 
\text{Houlsby-Amorosi-Rojas} & F /F _{0}=\left( 1+\widehat{\pi }%
_{s}^{2}/b\right) ^{1/2}, \\ 
\text{neo-hypoplasticity} & F /F _{0}=\left( 1+\widehat{\pi }%
_{s}^{2}/3\right) ^{1-\frac{n+\alpha }{2}},%
\end{array}
\label{GSH-other-Im}
\end{equation}%
with $F _{0}\equiv F \left( \widehat{\pi }_{s}=0\right) $ and 
\begin{equation}
\theta =\frac{2}{1+\sqrt{1-a\left( 2+a\right) {\xi }\widehat{\pi }
_{s}^{2}}}.  \label{theta-Im}
\end{equation}%
The shearing factor of  GSH explicitly shows the elastic instability that it contains: The dimensionless shear stress is restricted to  $0\le\widehat{\pi}_{s} \le\sqrt{a\left( 2+a\right) {\xi }}$. The root becomes imaginary at the right edge and instability sets
in. The other three models remain stable for all $\widehat{\pi }_{s}>0$. 

To account for the instability of a strictly static slope that is too steep, since there is no dynamics, one has only the energy to work with. A convex-concave transition -- very much in analogy to the van de Waals theory of the liquid-gas phase transition --  is an appropriate  candidate. Therefore, we prefer to use the elastic energy that contains these instabilities. One may alternatively use brute force and put in the end of elasticity by hand. But then an abrupt transition is forced onto the system, and one cannot account for any precursors. 
Circumstances are similar if one aims to account for the lack of elastic solutions at low densities, $\rho<\rho_{lp}$, relevant for the transition from elasto-plastic motion to fast flows. 
(We stress that the angle of repose is given by the critical value, see Sec.\ref{sapc}. If connected with  the appropriate dynamics, all four models can accounted for it.)
\begin{figure}[t] \begin{center}
\includegraphics[scale=0.6]{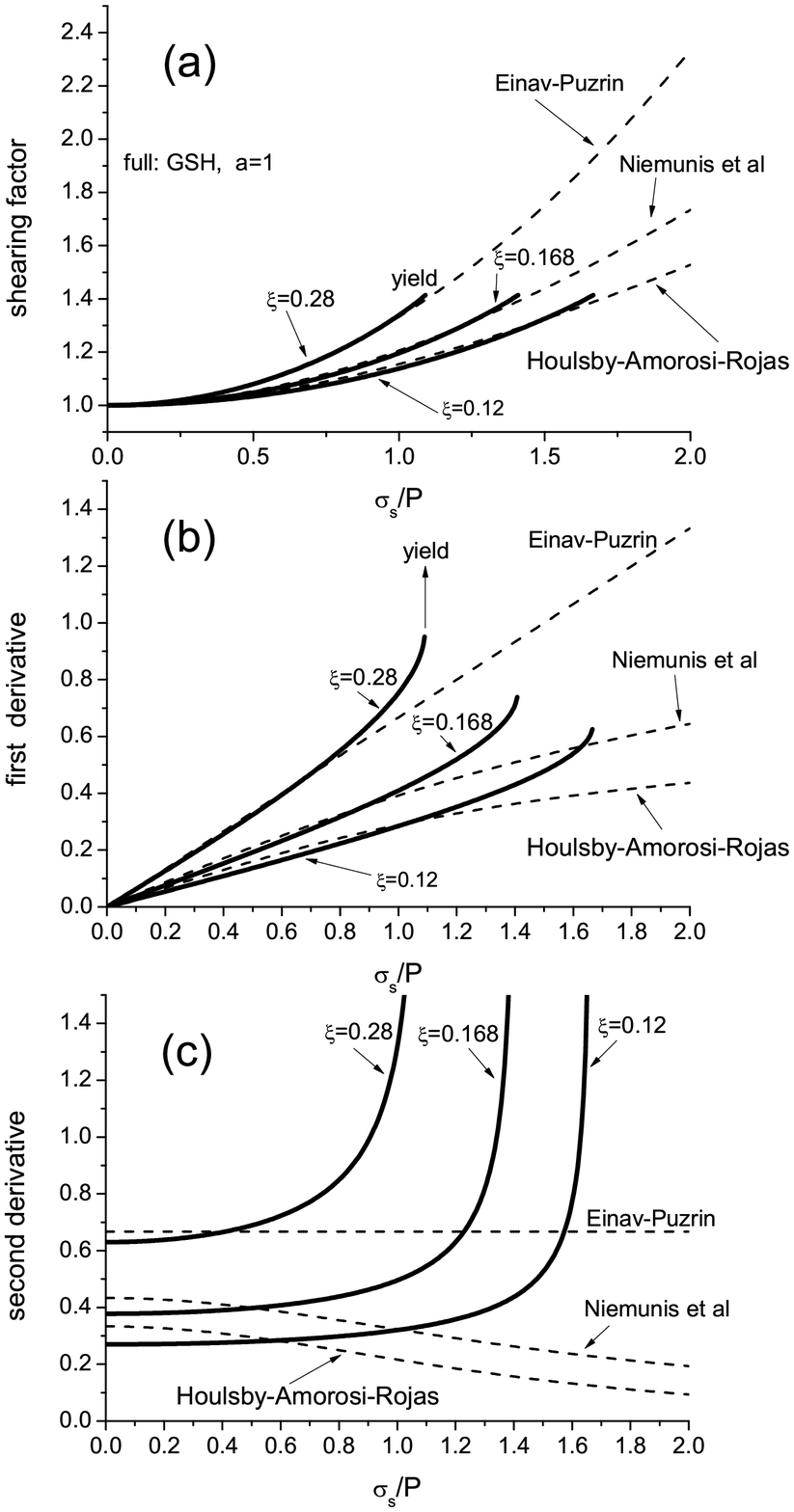}
\end{center}
\caption{\label{neohypo}} 
\caption {The shearing factors, as given in Eq.(\ref{GSH-other-Im}), and their
first, second derivatives.  }
\end{figure}

Taking $a=1$, all four models have essentially the same   dependence on $P_{\Delta }$, and the same ratio $w/\psi $. Then we only need to  compare the shearing factor $F $.  (We take  $b=3$ for Einav-Puzrin and Houlsby-Amorosi-Rajas.) Employing appropriate values for $\xi $, we can fit the shearing
factor of GSH to that of the other three, see  (a) of  Fig.(\ref{neohypo}), although the second derivative of GSH diverges,  see (c), laying bare the precursor of instabilities.

\subsubsection{The Third Strain Invariant\label{3inva}}
As stressed in the last section, an important advantage of the energy expression,  Eq.(\ref{2b-2}), is its built-in shear instability, Eq.(2b-3), that accounts for granular media's inability to sustain 
arbitrarily large {\em static shear stresses}. 
Humrickhouse observed that this instability occurs too early, as compared to the  {\em angle of stability} (the angle up to which a static layer of sand will remain at rest). And he proposed adding a term given by the third invariant of the elastic strain, $u_t^3=u_{ij}u_{jk}u_{ki}$, to increase the predicted angle of stability\cite{hum}, 
\begin{equation}\label{3inv}
w={\cal B}\sqrt\Delta\left[\frac25\Delta^2+\frac{\cal A}{\cal B}\left(u_s^2-\chi\frac{u_t^3}\Delta\right)\right].
\end{equation}
Unfortunately, considering only negative values of $\chi$, he came to the erroneous conclusion that this term does not yield any improvement. We subsequently found that it is possible to yield a realistic angle of stability of approximately $30^\circ$ with a positive $\chi$. What is more, incremental stress-strain relations and granular acoustics also support the same  $\chi$~\cite{3inv}.

\subsubsection{Proportional Paths and Barodesy\label{pep}}
Barodesy's starting point are the proportional paths as summed up by the Goldscheider rule (GR).
Taking {\sc p}$\varepsilon${\sc p} and {\sc p}$\sigma${\sc p} to denote proportional strain and stress paths, they are  
\begin{itemize}
\item A {\sc p}$\varepsilon${\sc p} starting from zero shear stress $\sigma_s=0$ is associated with a  {\sc p}$\sigma${\sc p}.
\item The same {\sc p}$\varepsilon${\sc p} starting from an arbitrary $\sigma_{ij}\not=0$
leads asymptotically to the same {\sc p}$\sigma${\sc p} obtained when starting at $\sigma_s=0$. 
\end{itemize}
Any constant strain rate $v_{ij}$ is a {\sc p}$\varepsilon${\sc p}: In the principal strain axes  $(\varepsilon_1,\varepsilon_2,\varepsilon_3)$, a constant $v_{ij}$ 
means the system moves with a constant rate along its direction, with $\varepsilon_1/\varepsilon_2=v_1/v_2,\, \varepsilon_2/\varepsilon_3=v_2/v_3$, independent of time. {\sc gr} states there is an associated stress path that is also a straight line in the principal stress space, that there are 
pairs of strain and stress paths. And if the initial stress value is not on the right line, it will converge onto it.  

To  understand GR using GSH, we need the stationary solution for an arbitrary {\sc p}$\varepsilon${\sc p}, ie. we need to generalize the stationary solution as given by Eq.(\ref{cs1},\ref{cs2}) to include $v_{\ell\ell}\not=0$. Using the superscript $^{ni}$ to mean non-isochoric, we have 
\begin{equation}
\label{eq74}
u_c^{ni}=u_c,\quad \frac{\Delta_c^{ni}}{u_c^{ni}}=\frac{\Delta_c}{u_c}+\frac{1-\alpha}{u_c\lambda_1f}\frac{v_{\ell\ell}}{v_s}\approx\frac{\Delta_c}{u_c},
\end{equation}
where the approximate sign holds because $v_{\ell\ell}\ll v_s$. We also note that the deviatoric stress, elastic strain and strain rate share the same direction,  $\hat\sigma_{ij}=\hat\pi_{ij}=-\hat u_{ij}=-\hat v_{ij}$, see Eqs.(\ref{sigma},\ref{2b-1},\ref{cs2}).  If the strain path is isochoric, $v_{\ell\ell}=0$, these are simply the critical state, with both the deviatoric elastic strain and stress being  constant. They are hence dots that remain stationary in their respective space, given by ${u_{ij}^{\ast }|_c}={u_c}(\rho)\,\hat v_{ij}$ and ${\sigma_{ij}^{\ast }}=- {\sigma_s}(\rho)\,{\hat v_{ij}}$. If however $v_{\ell\ell}\not=0$, with the density $\rho[t]$ changing accordingly, ${u_c}(\rho)$ and ${\sigma_s}(\rho)$ will also change, but not the direction $\hat v_{ij}$, making the dots walking down a straight line along $\hat v_{ij}$. 

This would indeed  be the first rule, except that GR states that it is the total (and not the deviatoric) stress that possesses a {\sc p}$\sigma${\sc p}.  
This is a slightly more involved point. With 
\begin{equation}
\sigma_{ij}=(1-\alpha)\pi_{ij}=[1-\alpha(\rho)]P_\Delta(\rho)[\delta_{ij}+(\pi_s/P_\Delta)\hat v_{ij}], 
\end{equation}
we again have a density dependent prefactor, $[1-\alpha(\rho)]P_\Delta(\rho)$,  multiplied by
\begin{equation}
\delta_{ij}+(\pi_s/P_\Delta)\hat v_{ij}\,\,\to\,\,\delta_{ij}+(\pi_c/P^c_\Delta)\hat v_{ij}.
\end{equation}
This factor is  a constant direction, giving rise to a  {\sc p}$\sigma${\sc p}, if we assume that the system is already in the critical state -- as discussed in Sec.\ref{Stationary Elastic Solution}), $\pi_s/P_\Delta$ does not depending on the density.
Otherwise, we need to rely on $\pi_s/P_\Delta$ being only weakly density dependent.

The second rule is much easier to understand. 
Given an initial elastic strain deviating from that prescribed by Eq~(\ref{eq74}),  $u^0_{ij}\not=u_c\hat v_{ij}$, $\Delta_0\not=\Delta_c$, Eqs~(\ref{eqU},\ref{eqD}) clearly state that any deviating component will relax toward zero, implying the elastic strain and the associated stress will converge onto the prescribed line.

\section{Experiments at Given Shear Stress\label{aging}} 

In this section,  we examine the ramification of GSH if, instead of the shear rate, the stress or elastic strain is being held constant.  As we shall see, 
taking  $\partial_tu_{ij}^*=0$ in Eq.(\ref{eqU}) results in a creep motion of the magnitude $fv_s/T_g=u_s/u_c$, such that an initial $T_g$ will co-relax with $v_s$, at the altered rate $R_T(1-\sigma_s/\sigma_c)$. Moreover, $T_g\propto v_s$ relax back to the equilibrium value $T_g,v_s=0$ only for $\sigma_s<\sigma_c$,  growing without bounds  for $\sigma_s>\sigma_c$, though a shear band solution is stable for  $\sigma_s>\sigma_c$. Note hypoplasticity is not applicable here. 

This behavior has implications for the angle of repose  $\varphi_{re}$ (for uniform stresses): As long as the shear stress is held below the critical one, $\sigma_s<\sigma_c$, the $T_g\propto v_s$-relaxation will run its course, and the system is in a static, mechanically stable state  afterwards. For $\sigma_s\geq\sigma_c$, however, the system does not come to a standstill. Therefore, defining $\varphi_{re}$ same as in Eq.(\ref{sb12}), but employing the critical rather than yield value, we have
\begin{equation}\label{dynfriAng}
\tan\varphi_{re}={\sigma_c}/\sqrt2\,P,\quad\text{with\,\,}\varphi_{re}<\varphi_{st}.
\end{equation}
The inequality holds because the critical state is an elastic solution, while $\varphi_{st}$ of Eq.(\ref{sb12}) is the angle at which all elastic solutions become unstable.

Stress-controlled experiments cannot be performed in a conventional triaxial apparatus possessing  stiff steel walls, because the correcting rates employed by the feedback loop to keep the stress constant are usually too strong. As a result, too much $T_g$ is excited that distorts its relaxation. The situation is then more one of consecutive constant rates, less of constant stress. Instead, one may employ a soft spring to couple the granular system with its driving device, to enable small-amplitude stress corrections without exciting much $T_g$.

Typical examples of stress-controlled experiments are creep (both uniform in space and exponentially decreasing), shear band, also flow on an inclined plane or in a rotating drum. The flow either comes to a halt, becomes jammed  at the angle of repose $\varphi_{re}$, or starts flowing from that jammed state, becomes fluidized at the  angle of stability $\varphi_{st}$. 
\subsection{Uniform Creep\label{TgRelaxation}}
For $T_g=0$, the system is static, $\sigma_s=$ const, and there 
is no dynamics, only a stable, static elastic solution. But if $T_g\not=0$ initially, both $T_g$ and the elastic strain will relax according to Eqs.(\ref{Tg2},\ref{eqU},\ref{eqD}), and with them also the stress $\sigma_s$. 
Maintaining a constant  $u_s$, or $\sigma_s$, therefore requires a compensating shear 
rate $v_s$, which is observed as creep.  
As we shall see, the characteristic time of $T_g$ is then $\propto(1-\sigma_s/\sigma_c)^{-1}$, hence rather long close to the critical stress. So the accumulated shear strain $\varepsilon_s(t)=\int v_s{\rm d}t$ can be expected to be large. 

In a recent experiment involving a fan submerged in sand, Nguyen et al.~\cite{aging} pushed the system 
to a certain shear stress at a given (and fairly fast) rate, producing an elevated $T_g$. Then, switching to maintaining the shear stress, they observed the accumulation of a large total strain  $\varepsilon_s(t)$ that appears to diverge logarithmically. They called it {\it creeping}.  

In the experiment, a very soft spring was used to couple
the fan and the motor. This fact we believe is essential why the
experiment turned out as observed. Usually, triaxial apparatus with stiff steel walls and a feedback loop is used to keep the stress constant. 
The correcting rates are strong, and much $T_g$ is excited that distorts its relaxation. The situation is hence more that of consecutive constant rates, less of constant stress. A soft spring coupling the granular system with its driving device has much lower correcting rates. 

Because of the fan,  the stress distribution in the experiment is rather nonuniform,  rendering a quantitative comparison to GSH difficult. Nevertheless, the experiment's new and structurally robust results are worth closer scrutiny. And we shall employ GSH assuming uniformity to yield a qualitative understanding.   Also, we shall assume it is the elastic shear strain $u_s$ that is being kept constant, not the shear stress $\sigma_s\propto\sqrt\Delta\,u_s$, as both cases will turn out to be rather similar.
The relevant equations are still Eqs~(\ref{Tg2},\ref{eqU},\ref{eqD}) (with  $T_a=0$). 

At the beginning, as the strain is being ramped 
up to $u_s$ employing the constant rate $v_1$, granular temperature acquires the initial value $T_0=fv_1$. Starting at the time $t=0$, the elastic strain $u_s$ is being held constant. Inserting $\partial_tu_s=0$ into Eq~(\ref{eqU}), we have
\begin{equation}\label{3b-7a}  
fv_s/T_g={u_s}/{u_c},
\end{equation}
with $v_s$ the rate needed to compensate the stress relaxation.
Inserting this into Eqs~(\ref{Tg2},\ref{eqD}), we arrive at
\begin{align}
\label{3b-8a} \partial_t\Delta=-\lambda_1T_g[\Delta-(u_s/u_c)^2\Delta_c],
\\\label{3b-9} 
\partial_t T_g= -R_T[(1-u_s^2/u_c^2)T_g^2-\xi_T^2T_g\nabla^2T_g],
\end{align}
with the effective $T_g$-relaxation rate reduced from $R_T$ to 
\begin{equation}
r_T\equiv R_T[1-u_s^2/u_c^2]. 
\end{equation}
Assuming uniformity, both equations may be solved analytically, if the coefficients are constant. Being functions of the density, they are if the density is. The stress $P(t),\sigma_s(t)$ will then change with time,  as will $\Delta(t)$. 
The first equation accounts for the exponential decay of $\Delta$, from both below and above, to the steady state value \begin{equation}\label{3b-11}
\Delta/\Delta_c=u_s^2/u_c^2.
\end{equation}
The relaxation is faster the more elevated $T_g$ is. 
Employing the initial condition $T_g=T_0$ at $t=0$, the solution to the second equation is
\begin{equation}\label{3b-10} 
T_g={T_0}/({1+{r_T}{T_0}t}).
\end{equation}
Because of Eq~(\ref{3b-7a}),  the same solution also holds for the shear rate, 
$v_s={v_0}/(1+ r_vv_0 t)$, with $v_0\equiv (u_s/u_c)T_0/f$, $r_v\equiv(fu_c/u_s){r_T}$,
implying a logarithmically divergent total shear strain,
\begin{equation}
\varepsilon_s-\varepsilon_0\equiv\int v_s{\rm d} t=\ln(1+r_vv_0t)/r_v.
\end{equation}
However, we note that $\varepsilon_s$ does not actually diverge, 
because as $T_g\to0$, it enters the quasi-elastic regime,  see Eq.(\ref{elatran}) below, 
where its relaxation becomes exponential. So even if creep is large close to $u_c$, it comes to a halt eventually, and the system is mechanically stable. 

Assuming a large $T_0$, $\Delta$ is quickly relaxed. 
Fixing $u_s$ is then equal to fixing the shear stress, $\sigma_s\propto\pi_s\propto u_s\sqrt{\Delta} =(u_s^2/u_c)\sqrt{\Delta_c} $. In addition, with
$\pi_s/\pi_c=\sqrt{\Delta}\,u_s/\sqrt{\Delta_c}\,u_c=(u_s/u_c)\sqrt{\Delta_c}\,u_s/\sqrt{\Delta_c}\,u_c=u_s^2/u_c^2$, one may rewrite the factor in $r_T$ as
\begin{equation}\label{3b-12} 
1-u_s^2/u_c^2=1-\pi_s/\pi_c=1-\sigma_s/\sigma_c.
\end{equation}
The $T_g$-relaxation is slower the closer  $\sigma_s$ is to $\sigma_c$, infinitely so  for  $\sigma_s=\sigma_c$. Then $u_s=u_c$, $\Delta=\Delta_c$,   with $T_g(t)=T_0=fv_1$ a constant that does not relax.  
This is indistinguishable from the rate-controlled critical state, which, clearly,  may be  maintained also at given stress. 

For $\sigma_s>\sigma_c$, the relaxation rate is negative, and $T_g\propto v_s$ will  grow, or explode, seemingly without bound, see the next two sections, each with a possibility of what happens next. 

Only a $T_g$ sufficiently large will explode,  destabilizing a static shear stress exceeding $\sigma_c$, an infinitesimal $T_g$ will not. This is because the critical stress diverges for $T_g\to0$ [as can be seen calculating the critical stress employing Eq.(\ref{elatran}) below], and the window between $\sigma_c$ and $\sigma_s^{yield}$ vanishes. Therefore, a static shear stress remains metastable for $T_g=0$, turning instable only at  the yield stress, $\sigma_s^{yield}/\sqrt2 P=\sqrt{{\cal A}/{\cal B}}$, as given in Eq.(\ref{sb12}).  

To summarize, as long as $\sigma_s<\sigma_c$, both the temperature and the shear flow will relax to zero, with the rate $r_T\equiv R_T[1-\sigma_s/\sigma_c]$. For  $\sigma_s=\sigma_c$, we have the rate-independent critical state. For $\sigma_s^{yield}>\sigma_s>\sigma_c$, the system is meta-stable, and easily perturbed into developing a shear band as given in the next section, \ref{nsb}. For $\sigma_s>\sigma_s^{yield}>\sigma_c$, the stress necessarily has the full form of Eq.(\ref{sigma}), containing both the gaseuous pressure, viscous stress, and also  the elastic part that is always in the critical state, see Sec.\ref{sapc}.


\subsection{Narrow Shear Bands\label{nsb}}

For $\pi_s^{yield}>\pi_s>\pi_c$, a stable, localized steady-state solution exists that we may identify as the shear band: Taking $\partial_tT_g=0$ in Eq.(\ref{3b-9}), the steady-state solution for $0\le x\le\xi_{sb}$ is
\begin{align}\nonumber
\nabla^2 T_g&=-T_g/\xi^2_{sb},\quad{\xi}^2_{sb}\equiv\xi_T^2/[\sigma_s/\sigma_c-1],
\\
 v_s/v_s^0&=T_g/T_g^0=\sin(\pi x/\xi_{sb}).\label{nsb2}
\end{align}			
The velocity difference across the band is $\Delta v=\int v_s{\rm d}x=\int v_s^0\sin(\pi x/\xi_{sb}){\rm d}x$, hence 
\begin{equation}
v_s^0 =\Delta v/(2\xi_{sb})=\sqrt{\sigma_s/\sigma_c}\,T_g^0/f.
\end{equation}
This is to be combined with $v_s\propto T_g\equiv0$ for $x\le0$ and $x\ge\xi_{sb}$.
We note that although $T_g$ and $v_s$ are continuous at $0, \xi_{sb}$, neither is differentiable there. Also, the density $\rho$ must be lower in the band, because only than can the shear stress $\sigma_s$ (that has to be uniform) be smaller than the critical stress  $\sigma_c(\rho)$ in the quiescent region, but larger in the band.

\subsection{Angle of Stability and Repose \label{sapc}}

If the system stays uniform at a shear stress exceeding the critical one, $\sigma_s>\sigma_c$, the variables $T_g\propto v_s$ will grow  with $r_T$, until the gaseous pressure and viscous stress become relevant, see Eq.(\ref{sigma}), while the elastic stress stays  in the critical state, 
\begin{align}\label{sigma2}
\sigma_{ij}=(1-\alpha)\pi^c_{ij}+P_T-\eta T_gv^*_{ij}, \quad T_g=fv_s.
\end{align}
This is the stress expression for fast dense flow, again a stable and uniform solution, though no longer rate-independent. We shall not dwell on it here, see~\cite{granRexp, PG2013} for results. 

On a tilted plane with a ongoing granular flow and a stress as given by Eq.(\ref{sigma2}), reducing the inclination angle will reduce both the shear stress and the shear rate $v_s$. And the flow will come to a stop, $T_g\propto v_s\to0$, if the stress drops below the critical one, as discussed in Sec.~\ref{TgRelaxation}. Hence   
the angle of repose (for uniform stress) is $\tan\varphi_{re}={\sigma_c}/\sqrt2\,P$, see Eq.(\ref{dynfriAng}. 

Given a static elastic state on a plane inclined by the angle $\varphi$, the angle of stability $\varphi_{st}$ (again for uniform stress) is reached when the energetic instability of Eqs.(\ref{2b-3}) is breached, and the static elastic state collapses. Therefore,  $\tan\varphi_{st}=\sigma_s^{yield}/\sqrt2 P=\sqrt{{\cal A}/{\cal B}}$, see Eq.(\ref{sb12}). 


\subsection{Wide Shear Bands and KCR\label{creep motion}}
Finally, we consider the coexistence between fast dense flow,  $T_g=fv_s$, and  one of static elastic stress, $T_g,v_s=0$. 
Some $T_g$ will diffuse into the  static region, giving rise to an exponentially decreasing creep, as observed by Komatsu et al~\cite{komatsu}, Crassous et al~\cite{crassous}.
 
We take this ``liquid-solid boundary"  to be at $x=0$, with $x>0$ being solid. In the fluid phase $x<0$, the shear rate is a constant, as is $T_g$, providing the boundary conditions for the solid part. Pressure $P$ and shear stress $\sigma_s$ are uniform, as is  $v_{\ell\ell}\equiv0$, but $\rho$ is discontinuous at $x=0$, though  constant otherwise.  (The liquid density needs to be lower than that of the solid one, such that the same stress exceeds the critical one in the liquid, but is below it in the solid.) The variable $T_g=fv_s$ varies  along $\hat x$.  These circumstances are similar to that of Sec~\ref{TgRelaxation}, though variation is in space rather than time. Given stationarity of  Eqs~(\ref{eqU},\ref{eqD}), we have
\begin{equation}\label{props}
\frac{\Delta}{{\Delta_c}}=\frac{u_s^2}{u_c^2}=\frac{f^2v_s^2}{T_g^2}=\frac{\pi_s}{\pi_c}=\frac{\sigma_s}{\sigma_c},
\end{equation}
see Eqs.(\ref{3b-7a},\ref{3b-11}). With $\Delta,u_s$ fixed, so are $P,\sigma_s$, where especially $P=P_c$ if $\sigma_s=\sigma_c$. Note that the above relations imply  $\Delta/{u_s}=({\Delta_c}/{u_c})({fv_s}/{T_g})$. And since $1/\mu\equiv P/\sigma_s$  increases monotonically with $\Delta/{u_s}$, see Eq.(\ref{2b-1}),  the friction $\mu$ decreases for increasing $fv_s/T_g$. 

The balance equation for $T_g$ with $\partial_tT_g=0$, see Eq.(\ref{3b-9}), yields an evanescent creep,
\begin{align}
\label{cr1}
\nabla^2T_g&=T_g/\xi^2_{cr},\quad \xi_{cr}^2\equiv\xi_T^2/[1-\sigma_s/\sigma_c],
\\\label{cr2}
v_s/v_s^{0}&= T_g/T_g^{0}= \exp(-x/\xi_{cr}).
\end{align}
($T_g^{0}$ and $v_s^{0}\sqrt{}$are the liquid values at $x=0$.) 
That the decay length $\xi_{cr}$ diverges for $\sigma_s=\sigma_c$ is not surprising, because the whole solid region turns critical then. 

Kamrin et al~\cite{kamrin,kamrin2}  proposed a constitutive relation (KCR) to account for the steady flow in the split-bottom cell~\cite{fenistein,fenisteinB}. A key ingredient of the theory is a new variable {\em fluidity}: $g\equiv v_s/\mu$, with  $\mu\equiv\sigma_s/P$ and $\mu_s\equiv\sigma_c/P_c$. It is taken to obey the equation
\begin{equation}
\xi^2_{cr}\nabla^2g=g-g_{loc},\quad \xi_{cr}\propto1/\sqrt{|\mu-\mu_s|}.
\end{equation}
As $g_{loc}=0$ for $\mu<\mu_s$, this relation is rather similar to Eq.(\ref{cr1}), with  $g$ assuming the role of $T_g$, and the decay length diverging at the critical stress.
For $\mu\geq\mu_s$, the system is fluid, and $g=g_{loc}$ essentially constant. Again, $T_g$ is constant in the liquid region. 

KCR is well capable of accounting for the steady flow in the split-bottom cell. This is fortunate,  because again, there is a symbiotic relation  relation between GSH and KCR, similar to that between GSH and hypoplasticity. But 
there are also three differences. 

First, KCR assumes $\sigma_s>\sigma_c$ in the liquid and  $\sigma_s<\sigma_c$ in the solid, with $\sigma_c$ a constant. This violates momentum conservation, as the stress is continuous across the solid-fluid interface (see also the discussion in Sec.\ref{nsb} and \ref{creep motion}).
Second, we have, for GSH, ${fv_s}/{T_g}=1$ in liquid, and ${fv_s}/{T_g}=\sqrt{\sigma_s/\sigma_c}$ in solid; yet $v_s/g=\mu\equiv \sigma_s/P$ on both sides for KCR. 
Third, Kamrin and Bouchbinder constructed a ``two-temperature continuous mechanics'' in a recent paper~\cite{ttcm}, defining a configuration temperature $\theta_c$ and one for the vibrational degrees of freedom $\theta_v$.  (Microscopic degrees of freedom are not considered. There would be three temperatures if one did.)  They hypothesize that the fluidity  $g$ may be related to   $\theta_c$. We disagree here. As discussed above,  $g$ is essentially $T_g$ of GSH, which is closer to  $\theta_v$.

\section{The Quasi-Elastic Regime\label{quasiEla}}

 {\it Quasi-static} is what many in soil mechanics call the rate-independent  elasto-plastic motion. The implication is: Being the slowest one possible, elasto-plastic motion stays as is however slow the rate.  We believe there is, at very small $T_g$ and very low rates -- possibly  $v_s\sim10^{-6}$/s, or $I\sim10^{-8}-10^{-9}$ (where $I\propto v_s/\sqrt P$ is the inertial number of the $\mu(I)$ rheology),  {\it a rate-dependent transition to another rate-independent regime.} If true, only the latter should be called quasi-static, or  {\it quasi-elastic}. The  reasons for our believe are the following six points [see also Sec.\ref{elasticLimit} and the list after Eq.(\ref{2b-1})]:
\begin{enumerate}

\item We have $T_g=0$ in static stress distributions, which is well accounted for by the fully elastic  Eqs.(\ref{sigmaela}). This should remain so for very slow stress changes.

\item The fact that cyclic load or ratcheting are  less dissipative than hypoplasticity predicts~\cite{Herle} may well be interpreted as the onset of a transition to quasi-elasticity. One does not need to invoke intergranular strain here.  

\item Quasi-static motion --  a  consecutive visit of static, equilibrium states --   should occur at $T_g=0$, and be reversible. Yet elasto-plastic motion occurs at elevated $T_g$, is strongly dissipative and irreversible. Reactive and dissipative terms are comparable in size, being exactly equal in the critical state. 
(This is obvious only within the context of GSH, but one can identify the reactive and dissipative terms in  hypoplasticity by comparing it to GSH. Then the same holds true there.)
 
\item Rate-independence is a hallmark of quasi-static motion. Yet the  rate-independence of elasto-plastic motion is (as discussed in Sec.\ref{external perturbation}) easily destroyed. 

\item $T_g$  is essentially zero in elastic waves, since there is not enough time to excite appreciable amount of it. Waves are well accounted for by the elastic Eqs.(\ref{sigmaela}), but not by hypoplasticity -- as they should if there were only the elasto-plastic regime. Employing hypoplasticity, one finds that waves are  always over-damped: { Starting from momentum conservation, $\partial_t(\rho v_i)=-\nabla_j\sigma_{ij}$, or $\,\,\partial^2_t(\rho v_i)=-\nabla_j\partial_t\sigma_{ij}$, one inserts Eq.(\ref{3b-1}) to find the term $\propto H_{ijk\ell}$ yielding the velocity and the one $\propto\Lambda_{ij}$  damping. Both terms are of the same order in $q$, and given the values of the two tensors appropriate for elasto-plastic motion, they are also comparable in size, implying that elastic waves are always over-damped. (In solids, the damping term is an order in the wave vector $q$ higher than the velocity term, rendering it much smaller  for long enough wave lengths. This is the reason elastic waves propagate there.)}.
In fact, elastic waves are a fast yet quasi-static phenomenon, same as incremental stress-strain relations. 

 \item Happily and perhaps most convincingly, Peng Zhen et al.  have most recently observed indications of a rate-dependent transition away from the elasto-plastic  regime~\cite{QuasiEla}. 

\end{enumerate}

\subsection{ Strain Increments  versus Slow Rates}
Faced with the quasi-elastic response of sound waves and ratcheting, a frequent suggestion is to take a small strain  increment to be elastic and free of dissipation, but a large one as elasto-plastic and dissipative. Unfortunately, this is not  compatible with the notion of  quasi-static motion, a consecutive visit of equilibrium, static states: Starting from an arbitrary static state of given stress, applying a small incremental strain that is elastic, the system ends up in another static state, with a slightly higher stress. This second state, as static as the first, is again a valid starting point -- any static state is. And the next small increment must again be elastic. Many consecutive small increments yield a large change in strain, and if the small ones are not dissipative, neither can their sum be. 

In GSH , it is the strain rate, not the strain amplitude, that decides whether the response is elastic or elasto-plastic.  Small strain increments produced with
a brief puls of shear rate will produce an elastic response, if
$T_g$ does not have time to get to a sufficiently high value.

\subsection{How to Observe Quasi-Elasticity}
There are at least two ways to observe the elusive quasi-elastic behavior, both by fixing the stress rate and keeping a low $T_g$: Note that a given stress rate may be associated with either of two shear rates, a high, elasto-plastic one at elevated $T_g$ and a low, elastic one at vanishing $T_g$.  
The first method is to slowly incline a plane supporting a layer of grains. In such a situation, the shear rate remains very small, and the system starts flowing only when the yield stress is breached. 

A second method is to insert a {soft spring} between the granular
medium and the driving device. If the spring is softer by a large factor $a$ than the granular medium (which is itself rather soft), it will  serve as a ``stress reservoir,'' and absorb most of the displacement, leaving the granular medium deforming at a rate smaller by the same factor $a$. (Employing a feedback loop in a triaxial apparatus with stiff steel walls to maintain a stress rate will not usually work, because the correcting motion typically has strain rates that are so high, that the system slips into an elasto-plastic motion quickly.)

\subsection{Transition to Quasi-Elasticity}
To account for the transition to quasi-elasticity, we need to specify how, in Eqs.(\ref{2c-7},\ref{2c-8},\ref{sigma}),   $\lambda T_g,\lambda_1T_g,\alpha,\alpha_1$ vanish.  
There are scant data to rely on, but one possibility would be
\begin{equation}\label{alphaEla}
\alpha/\alpha^0=\alpha_1/\alpha^0_1= T_g/(T_g+T_0).
\end{equation} 
For $T_g\gg T_0$, we have $\alpha=\alpha^0,\,\, \alpha_1=\alpha^0_1$, ensuring rate-independence for the elassto-plastic regime; for $T_g\ll T_0$, we have $\alpha, \alpha_1\propto T_g$ vanishing with $T_g$. 

Instead of taking $\lambda,\lambda_1\to0$ with $T_g$, we choose to generalize $T_g^2=f^{2}v_s^2$ to
\begin{equation}\label{elatran}
T_g^2=f^2v_s^2/(1+T_1/T_g),
\end{equation} 
because this was a natural result arising during the derivation of GSH, see~\cite{granR4,granRexp}. For  $T_g\ll T_1$, we have $T_g=(f^2/T_1)\,v_s^2$ quadratically small in $v_s$, and with it also the relaxation rates $\propto T_g$. The ratio,  in Eqs.(\ref{2c-7},\ref{2c-8}), between the irreversible, plastic term ($\propto v_s^2$) and the reversible, elastic term ($\propto v_s$) diminishes with $v_s\to0$, as has been observed in DEM~\cite{AH}.

\section{Conclusion}
Taking the stress increment as a function of density, strain rate and the stress itself, the Eq.(1) this paper starts from to set up a constitutive model, is an appropriate and physically sound way to come to terms with the complexity of triaxial results.
A rather broader view of granular behavior are  captured by introducing a pair of internal state variables, the elastic strain field $u_{ij}$ and the granular temperature $T_g$, with the first accounting for the coarse-grained deformation of the grains, and the latter accounting for their quickly changing elastic and kinetic energy. The theory that does it is called GSH, and the phenomena accounted for include static stress distributions, creep, shear band, angle of repose and stability, also the critical state under tapping, all in addition to the trixial results.



\begin{thebibliography}{99} 
\bibitem{schofield}
P.~Wroth A.~Schofield.
\newblock {\em Critical State Soil Mechanics}.
\newblock McGraw-Hill, London, 1968.

\bibitem{wood1990}
D.~M. Wood.
\newblock {\em Soil Behaviour and Critical State Soil Mechanics}.
\newblock Cambridge University Press, 1990.

\bibitem{kolymbas1}
D.~Kolymbas.
\newblock {\em Introduction to Hypoplasticity}.
\newblock Balkema, Rotterdam, 2000.

\bibitem{kolymbas2}
W.~Wu and D.~Kolymbas.
\newblock {\em Constitutive Modelling of Granular Materials}.
\newblock Springer, Berlin, 2000.
\bibitem{nedderman}
R.M. Nedderman.
\newblock {\em Statics and Kinematics of Granular Materials}.
\newblock Cambridge University Press, 1992.

\bibitem{gudehus2010}
G.~Gudehus.
\newblock {\em Physical Soil Mechanics}.
\newblock Springer SPIN, 2010.

\bibitem{barodesy} Kolymbas D. Barodesy: a new constitutive frame for soils. Geotechnique Letters 2, 17–23,  (2012), http://dx.doi.org/10.1680/geolett.12.00004; 
Barodesy: A new hypoplastic approach. International Journal for Numerical and Analytical Methods in Geomechanics (2011). doi:10.1002/nag.1051;
Sand as an archetypical natural solid. In Mechanics of Natural Solids, Kolymbas D, Viggiani G (eds.). Springer: Berlin, (2009); 1–26; 

\bibitem{h^2} I.~Einav. 
\newblock The unification of hypo-plastic and elasto-plastic theories.
\newblock{\it International Journal of Solid and Structure}, {\bf 49}(2012) 1305-1315


\bibitem{GSH&Barodesy} Yimin Jiang, and Mario Liu. Proportional Path, Barodesy, and Granular Solid Hydrodynamics.  {\it Granular Matter} {\bf 15}, 237 (2013).

\bibitem{wu2}Wei Wu. {\em On high-order hypoplastic models for granular materials}. Journal of Engineering Mathematics {\bf 56}: 23–34 (2006) 

\bibitem{wu1}Tejchman, J. and Wu, W. {\it FE-investigations of micro-polar boundary conditions along interface between soil and structure}, Granular Matter, {\bf 12}, 399 (2010)

\bibitem{deGennes}
P.G. de~Gennes and J.~Prost.
\newblock {\em The Physics of Liquid Crystals}.
\newblock Clarendon Press, Oxford, 1993.


\bibitem{Herle}
A. Niemunis and I. Herle
{\it Hypoplastic model for cohesionless soils with elastic strain range}, 
MECHANICS OF COHESIVE-FRICTIONAL MATERIALS, {\bf 2}, 279 (1997)

\bibitem{Khal}
I.~M. Khalatnikov.
\newblock {\em Introduction to the Theory of Superfluidity}.
\newblock Benjamin, New York, 1965.

\bibitem{LL6}
L.~D. Landau and E.~M. Lifshitz.
\newblock {\em Fluid Mechanics}.
\newblock Butterworth-Heinemann, 1987.

\bibitem{granL3}
Y.M.~Jiang and M.~Liu.
\newblock From elasticity to hypoplasticity: Dynamics of granular solids.
\newblock {\em Phys. Rev. Lett.}, 99(10):105501, 2007.

\bibitem{granR2}
Y.M.~Jiang and M.~Liu.
\newblock Granular solid hydrodynamics.
\newblock {\em Granular Matter}, 11:139, May 2009.
Free download: http://www.springerlink.com/content/a8016874j8868u8r/fulltext 

\bibitem{granR3}
Y.M.~Jiang and M.~Liu.
\newblock The physics of granular mechanics.
\newblock In D.~Kolymbas and G.~Viggiani, editors, {\em Mechanics of Natural
  Solids}, pages 27--46. Springer, 2009.

\bibitem{granRgudehus}
G.~Gudehus, Y.M. Jiang, and M.~Liu.
\newblock Seismo- and thermodynnamics of granular solids.
\newblock {\em Granular Matter}, 1304:319--340, 2011.


\bibitem{GGas}Y.P. Chen, M.Y. Hou, Y.M. Jiang, and M. Liu
Hydrodynamics of granular gases with a two-peak distribution
{\it Phys. Rev.} {\bf E88}, 052204 (2013)

\bibitem{granR4}  
  Y.M.~Jiang and M.~Liu.
  \newblock Granular Solid Hydrodynamics  (GSH): a broad-ranged macroscopic theory of granular media.
\newblock {\em Acta Mech.}, {\bf 225}, 2363 (2014)



\bibitem{granRexp}
  Y.M.~Jiang and M.~Liu. {\it Applying GSH to a wide range of experiments
in granular media.} Eur. Phys. J. E (2015) 38:15


\bibitem{Fang1}
Chung Fang, Yongqi Wang, Kolumban Hutter. 
{\it A unified evolution equation for the Cauchy stress
tensor of an isotropic elasto-visco-plastic material}
Continuum Mech. Thermodyn. (2008) 19: 423–440
DOI 10.1007/s00161-007-0062-9

\bibitem{Fang2}
Chung Fang, Cheng-Hsien Lee. 
{\it A unified evolution equation for the Cauchy stress tensor
of an isotropic elasto-visco-plastic material;
II. Normal stress difference in a viscometric flow, and an unsteady flow
with a moving boundary}
Continuum Mech. Thermodyn. (2008) 19: 441–455
DOI 10.1007/s00161-007-0063-8

\bibitem{QuasiEla} Yan Xi-Ping, Peng Zheng, He Fei-Fei, Jiang Yi-Min. 
{\it Measurements of Shear Elasticity of Granular Solids},  to be published in 
Acta Phys. Sin.

\bibitem{Fang3}
Chung Fang. 
{\it A k-$\varepsilon$ turbulence closure model of an isothermal dry
granular dense matter}, to be published in
Continuum Mech. Thermodyn. DOI 10.1007/s00161-015-0454-1

\bibitem{polymer-1} H. Temmen, H. Pleiner, M. Liu and H.R. Brand, {\it
Convective Nonlinearity in Non-Newtonian Fluids,} Phys. Rev. Lett. {\bf
84}, 3228 (2000). 

\bibitem{polymer-2}H. Temmen, H. Pleiner, M. Liu and H.R.
Brand,{\it Temmen et al. reply}, Phys. Rev. Lett. {\bf 86}, 745 (2001).

\bibitem{polymer-3} H. Pleiner, M. Liu and H.R. Brand, {\it Nonlinear Fluid
Dynamics Description of non-Newtonian Fluids}, {Rheologica Acta} {\bf 43},
502 (2004). 

\bibitem{polymer-4}Oliver M\"uller, Mario Liu, Harald Pleiner, and Helmut R. Brand;
{\it Transient elasticity and polymeric ﬂuids: Small-amplitude deformations};
Phys.Rev. {\bf E 93}, 023113 (2016); and 
{\it Transient elasticity and the rheology of polymeric ﬂuids with large amplitude deformations};
Phys.Rev. {\bf E 93}, 023114 (2016).



\bibitem{ge1}
D.~O. Krimer, M.~Pfitzner, K.~Br\"auer, Y.M.~Jiang, and M.~Liu.
\newblock Granular elasticity: General considerations and the stress dip in
  sand piles.
\newblock {\em Phys. Rev. E)}, 74(6):061310, 2006.

\bibitem{ge2}
K.~Br\"auer, M.~Pfitzner, D.~O. Krimer, M.~Mayer, Y.M.~Jiang, and M.~Liu.
\newblock Granular elasticity: Stress distributions in silos and under point
  loads.
\newblock {\em Phys. Rev. E (Statistical, Nonlinear, and Soft Matter Physics)},
  74(6):061311, 2006.

\bibitem{granR1} Y.M. Jiang, M. Liu. Eur. A brief review of granular elasticity. Phys. J. \textbf{E~22,} 255 (2007).

\bibitem{KJ}  
R.~Kuwano and R.~J. Jardine.
\newblock On the applicability of cross-anisotropic elasticity to granular
  materials at very small strains.
\newblock {\em Geotechnique}, 52(10):727--749, Dec 2002.


\bibitem{AH}  F. Alonso-Marroquin, H.J. Herrmann,
Ratcheting of Granular Materials. Phys. Rev. Lett. \textbf{92}, 054301 (2004). 

\bibitem{incre}
Y.M.~Jiang and M.~Liu.
\newblock Incremental stress-strain relation from granular elasticity:
  Comparison to experiments.
\newblock {\em Phys. Rev. E (Statistical, Nonlinear, and Soft Matter Physics)},
  77(2):021306, 2008.
  

\bibitem{ge4}
M.~Mayer and M.~Liu.
\newblock Propagation of elastic waves in granular solid hydrodynamics.
\newblock {\em Phys. Rev. E}, 82:042301, 2010.

\bibitem{jia2009}
Y.~Khidas and X.~Jia.
\newblock Anisotropic nonlinear elasticity in a spherical-bead pack: Influence
  of the fabric anisotropy.
\newblock {\em Phys. Rev. E}, 81:021303, Feb. 2010.

\bibitem{3inv} Y.M. Jiang, H.P. Zheng, Z. Peng, L.P. Fu, S.X. Song, Q.C. Sun, M. Mayer, and M. Liu, \newblock Expression for the granular elastic energy.
\newblock  Phys. Rev. E {\bf 85}, 051304 (2012)    


\bibitem{hardin}
B.O. Hardin and F.E. Richart.
\newblock Elastic wave velocities in granular soils.
\newblock {\em J. Soil Mech. Found. Div. ASCE}, 89: SM1:33--65, 1963.


\bibitem{Sun} Qicheng Sun, Energy fluctuations at particle scale, preprint, (2015); 
Qicheng Sun, Granular structure and the nonequilibrium thermodynamics, Acta Phys. Sin. Vol. 64, No. 7 (2015) 076101; 
Qicheng Sun,  Feng Jin,  and Gordon G. D. Zhou, Energy characteristics of simple shear granular flows, Granular Matter (2013) 15:119–128; 
Q. Sun, S. Song, J. Liu, M. Fei, and F. Jin, Granular materials: Bridging damaged solids and turbulent fluids, THEORETICAL \& APPLIED MECHANICS LETTERS 3, 021008 (2013)

\bibitem{PG2013}
Yimin Jiang and Mario Liu, 
AIP Conf. Proc. 1542, 52 (2013); {\it Stress- and rate-controlled granular rheology,} http://dx.doi.org/10.1063/1.4811867

\bibitem{wichtmann} T. Wichtmann, Schriftreihe Inst. Grundbau u. Boden\-me\-cha\-nik, Univ. Bochum, Heft~38, (2005), Fig~4.17.
\bibitem{thornton} C. Thorn\-ton, S.J. Antony, {\it Phil.Trans.R.Soc.A: Mathematical, Physical and Engineering Sciences}, 
{\bf 356}, 
No. 1747, Mechanics of Granular Materials in Engineering and Earth Sciences (Nov. 15, 1998), 2763-2782 (1998).
\bibitem{vHecke2011}
J.A. Dijksman, G.H. Wortel, L.T.H. van Dellen,
O. Dauchot, and M. van Hecke. 
\newblock Jamming, yielding, and rheology of weakly vibrated granular media.
\newblock {\em Phys. Rev. Lett.}, {\bf 107},
108303(2011).
\bibitem{StressDip} D. Krimer, S. Mahle and M. Liu, Dip of the Granular Shear Stress, {\it Phys. Rev.} {\bf E86}, 061312 (2012)
\bibitem{P&G2009}  
  Y.M.~Jiang and M.~Liu.
  \newblock GSH, or Granular Solid Hydrodynamics:
on the Analogy between Sand and Polymers.
\newblock {\em AIP Conf. Proc.} 7/1/2009, Vol. 1145 Issue 1, p1096.

\bibitem{Nie3}H. A. Rondon, T. Wichtmann , Th. Triantafyllidis, A. Lizcano. 
Hypoplastic material constants for a well-graded granular material for base and subbase layers of flexible pavements. {\it Acta Geotechnica}, Vol. 1, No. 2, pp. 113-126, 2007


\bibitem{99} Andrzej Niemunis, Carlos E. Grandas Tavera, and Torsten Wichtmann:
\newblock{\em Peak stress obliquity in drained and undrained
sands. Simulations with neohypoplasticity,} T. Triantafyllidis (ed.), Holistic Simulation of Geotechnical
Installation Processes, Lecture Notes in Applied and Computational
Mechanics 80, DOI 10.1007/978-3-319-23159-4-5, Springer (2016)

 \bibitem{hum}
Paul W. Humrickhouse, PhD thesis, University of Wisconsin–Madison (2009);  P. W. Humrickhouse, J. P. Sharpe, and M. L. Corradini, Comparison of hyperelastic models for granular materials. Phys. Rev. E 81, 011303 (2010). 
\bibitem{granL1} Y. Jiang and M. Liu. Granular Elasticity without the Coulomb Condition. Phys. Rev. Lett. 91, 144301 (2003).

\bibitem{HousbyA}G. T. Houlsby, A. Amorosi, and E. Rojas, “Elastic moduli of soils dependent on pressure: a hyperelastic
formulation,” Geotechnique, vol. 55, no. 5, pp. 383–392, 2005.

\bibitem{EinavP} I. Einav and A. M. Puzrin, “Pressure-dependent elasticity and energy conservation in elastoplastic
models for soils,” Journal of Geotechnical and Geoenvironmental Engineering, vol. 130, no. 1, pp. 81–
92, 2004.

\bibitem{aging}
Van~Bau Nguyen, Thierry Darnige, Ary Bruand, and Eric Clement.
\newblock Creep and fluidity of a real granular packing near jamming.
\newblock {\em Phys. Rev. Lett}, 107:138303, 2011.


\bibitem{komatsu}
T.S. Komatsu, S.~Inagaki, N.~Nakagawa, and S.~Nasuno.
\newblock Creep motion in a granular pile exhibiting steady surface flow.
\newblock {\em Phys. Rev. Lett.}, 86:1757�1760, 2001.

\bibitem{crassous}
J~Crassous, J-F Metayer, P~Richard, and C.~Laroche.
\newblock Experimental study of a creeping granular flow at very low velocity.
\newblock {\em J. Stat. Mech.}, 2008:P03009, 2008.


\bibitem{kamrin}D.L. Henann and K. Kamrin. A predictive, size-dependent continuum model for dense granular flows. {\it Proceedings of  the National  Academy of  Sciences},{\bf 110}, 6730 (2012). http://www.pnas.org/content/110/17/6730.full. 

\bibitem{kamrin2} K. Kamrin and G. Koval. Nonlocal Constitutive Relation for Steady Granular Flow. {\it Phys.Rev.Lett.} {\bf108},178301 (2012)

\bibitem{fenistein} D. Fenistein, J.W. van de Meent, M.van Hecke. Kinematics: Wide shear zones in granular bulk flow. {\it Nature}, {\bf 425} 6955 (2003). 

\bibitem{fenisteinB} D. Fenistein, J.W. van de Meent, M.van Hecke. Core Precession and Global Modes in Granular Bulk Flow. {\it Phys.Rev.Lett.} {\bf 96}, 118001 (2004); {\bf 96}, 038001 (2006). 

\bibitem{ttcm}Ken Kamrin and Eran Bouchbinder.   {\it Journal of the Mechanics and Physics of Solids,} {\bf 73} 269–288  (2014). {\it Two-temperature continuum thermomechanics of deforming amorphous solids.}



\end{thebibliography}
\end{document}